\documentclass[twocolumn,preprintnumbers,superscriptaddress]{revtex4-1}
\usepackage{amssymb}
\usepackage{amsmath}
\usepackage{amsfonts}
\usepackage{graphicx}
\usepackage{dcolumn}
\usepackage{bm}
\usepackage[dvipsnames]{xcolor}

\newcommand{\be}{\begin{equation}}
\newcommand{\ee}{\end{equation}}
\newcommand{\bea}{\begin{eqnarray}}
\newcommand{\eea}{\end{eqnarray}}
\newcommand{\bes}{\begin{subequations}\begin{eqnarray}}
\newcommand{\ees}{\end{eqnarray}\end{subequations}}
\newcommand{\non}{\nonumber}
\newcommand{\sugi}[1]{{\bf (SS: #1)}} 
\newcommand{\real}{{\rm Re}}
\newcommand{\img}{{\rm Im}}

\begin{document}

\title{Resonantly enhanced polariton wave mixing and Floquet parametric instability}
\author{Sho Sugiura}
\affiliation{Department of Physics, Harvard University, Cambridge, Massachusetts 02138, USA}
\author{Eugene A. Demler}
\affiliation{Department of Physics, Harvard University, Cambridge, Massachusetts 02138, USA}
\author{Mikhail Lukin}
\affiliation{Department of Physics, Harvard University, Cambridge, Massachusetts 02138, USA}
\author{Daniel Podolsky}
\affiliation{Department of Physics, Technion, Haifa 32000, Israel}
\date{\today }

\begin{abstract}
We introduce a new theoretical approach for analyzing pump and probe experiments in non-linear acousto-optic systems.
In our approach, the effect of coherently pumped polaritons 
is modeled as providing time-periodic modulation of the system parameters. Within this framework, propagation of the probe pulse is described 
by the Floquet version of Maxwell's equations and leads to such phenomena as frequency mixing and resonant parametric production of polariton pairs. We analyze light reflection from a slab of insulating material with a strongly excited phonon-polariton mode and obtain analytic expressions for the frequency-dependent reflection coefficient for the probe pulse. Our results are in agreement with recent experiments by  Cartella et al. \cite{Cartella12148} which demonstrated light amplification in resonantly excited SiC insulator.
We show that, beyond a critical pumping strength, such systems should exhibit Floquet parametric instability, which corresponds to resonant  scattering of the pump polaritons 
into pairs of finite momentum polaritons. We find that the parametric instability should be achievable in SiC using current experimental techniques and discuss its signatures, including the non-analytic frequency dependence of the reflection coefficient and the probe pulse afterglow. We discuss possible applications of the parametric instability phenomenon and suggest that similar types of instabilities can be present in other photoexcited non-linear systems.
\end{abstract}

\maketitle

\section{Introduction}

\subsection{Motivation and overview}

Control of light by collective excitations in solids, such as phonons, opens exciting possibilities for realizing materials and devices with new functionalities \cite{LeitenStorferPRL, LeitenstorferPRB}. Recent experiments
by Cartella et al. \cite{Cartella12148} demonstrated terahertz optical amplification in SiC insulator following a strong mid-IR pump that resonantly excited the SiC stretching mode. Motivated 
by these experiments we study the problem of light reflection from a slab of insulating material with a strongly excited polariton mode. We analyze this system using 
the Floquet formalism, in which polariton oscillations provide periodic time modulation of the material properties. We present an analytic solution of the Fresnel light reflection problem in which we include mixing of two frequencies analogous to mixing between signal and idler modes in  parametric amplifiers. We demonstrate discontinuous dependence of the reflection coefficient on the incoming photon frequency, which we
attribute to the existence of unstable polariton modes. We interpret these instabilities as resonantly enhanced polariton parametric wave mixing.

\begin{figure}
		\includegraphics[width= 0.4 \textwidth]{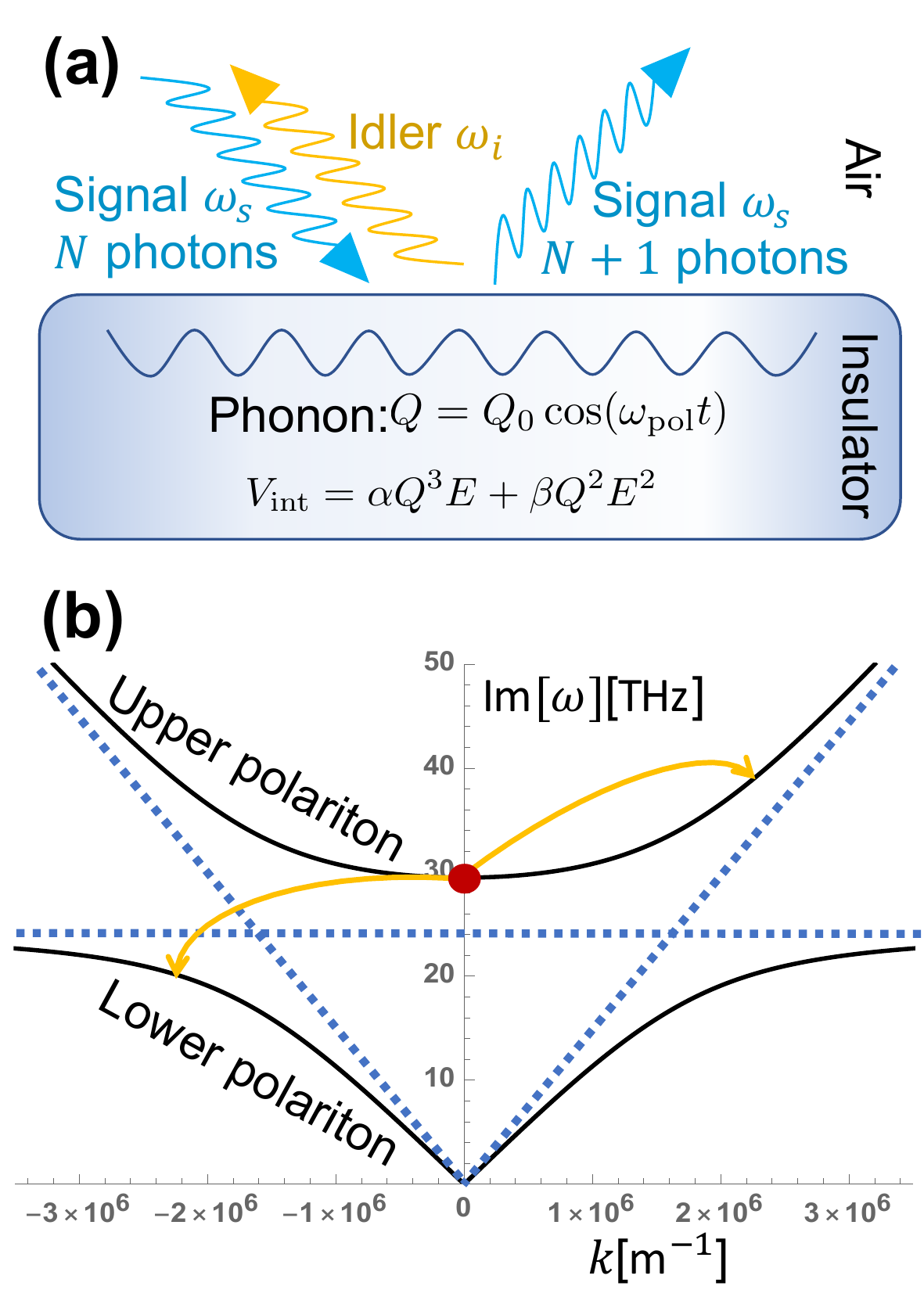}
		\caption{(a) Schematic figure of light amplification by a coherently oscillating polariton. Oscillating phonons result in parametric generation of pairs of photons that satisfy the condition $\omega_{\rm s}+\omega_{\rm i}=2\omega_{\rm p}$. Probe pulse results in Bose enhancement of the process in which one of the parametrically excited photons matches the incoming photon. Hence we find not only amplification of light at the incident frequency but also generation of light at the complimentary frequency. 
		(b)Floquet parametric instability as resonant scattering of polaritons. Dotted blue lines show schematically the bare phonons and photon dispersions. Solid black lines show the upper and lower polariton modes resulting from the hybridization of the photon and phonon modes. Polariton condensate at the bottom of the upper polariton band is generated by the pump pulse. Interaction between polaritons arising from non-linearities lead to an elastic scattering process that conserves both energy and momentum. Energies of out-coupled polaritons correspond to modes exhibiting Floquet parametric instability that we discuss in Sec.~\ref{Sec eigenmodes}
		}
		\label{fig schematic}
\end{figure}

A simple physical picture of light amplification by a coherently oscillating polariton emerging from coupling of light to phonons is shown in Fig.~\ref{fig schematic}a. Non-linear coupling between an IR active phonon $Q$ and electric field $E$ leads to the interaction term of the
form $Q^2\,E^2$. When the polariton mode is strongly excited by the pump pulse, we find coherent oscillations of the phonon field $\langle \, Q^2(t) \, \rangle = A \cos^2 \omega_{\rm pol} t$, which can be interpreted as
Floquet driving of the material. Such driving can result in the production of pairs of photons at frequencies  that add up to twice the drive frequency $\omega_1+\omega_2 = 2  \omega_{\rm pol}$.
This is analogous to four wave mixing in non-linear optics, with two of the waves corresponding to the collective mode of the material. When there is a probe pulse at frequency $\omega_{\rm s}$,
we find not only an increase in the number of photons at the same frequency upon reflection, but also generation of the complementary idler photons with frequency $\omega_{\rm i} = 2 \omega_{\rm pol} - \omega_{\rm s}$. 
If we neglect feedback of the signal pulse on the coherently oscillating polariton mode, we can analyze the light amplification effect by solving Maxwell's equations using the two-mode mixing Floquet formalism. 
In this case, reflection and transmission coefficients are described by the Floquet eigenmodes in which the components of the signal and idler frequencies are mixed. We note that nonlinearity arising from hybridization of a pair of photons with
a pair of polaritons is expected to be considerably stronger than the usual optical four wave mixing since it comes from a resonant process.

In principle, in Floquet systems one has mixing between more than two frequencies. We discuss generalization of our analysis beyond the two mode approximation and show that it does not change results appreciably. 
Moreover, the Floquet formalism also allows us to study unstable polariton modes induced by coherent oscillations of the phonon field. We consider the light reflection coefficient as a function of complex frequency and look for its poles in the complex plane. Our analysis treats the weak probe pulse only to linear order, hence, we can apply the standard linear response argument that poles of the response functions correspond to collective modes in the system.  Collective modes with positive imaginary part imply instabilities, corresponding to exponential growth in time of small initial fluctuations, which will ultimately be limited by non-linear terms. In the case of static gain media, the analysis of unstable polariton modes through poles of the reflection coefficients has been discussed in Refs. \cite{Johannes2006, Dorofeenko2012}.

A simple physical picture of the Floquet parametric instability comes from the perspective of resonant polariton wave mixing. Linear coupling between light and phonon $Q$ leads to formation
of upper and lower polariton modes, as shown in Fig.~\ref{fig schematic}b. An initial pump pulse creates a condensate of upper polaritons at zero momentum. Non-linear interactions 
between phonons and the electric field, such as $Q^4$, $Q^3 E$ and $Q^2\,E^2$, result in scattering of polaritons. In particular, there is a process in which two polariton modes scatter in a way that conserves energy and momentum (see Fig.~\ref{fig schematic}b). As we discuss below, this argument correctly predicts frequencies of the unstable modes, which are naturally connected by the complementarity 
condition $\omega_1+\omega_2 = 2  \omega_{\rm pol}$\cite{carusotto2013}. 

Our results provide a simple physical interpretation of light amplification observed in recent experiments by Cartella et. al. \cite{Cartella12148}. 
Moreover, we argue that Floquet parametric instability instability could already be present in these experiments since the critical pumping field strength is smaller than the one 
used by the Hamburg group. 
Our approach utilizing the Floquet formalism is applicable to a broad class of systems, and thus, it provides a general framework for investigating instabilities in pumped systems.
Finally, our analysis reveals a general mechanism that could be useful for realization of practical sources of THz radiation, relevant for many potential applications. 

\subsection{Relation to earlier work on optical phase conjugation}

Before proceeding, it is useful to put our results in the context of earlier work on optical phase conjugation and optical phase conjugation phenomena \cite{Fisher,Zeldovich}. 
While phase conjugation phenomena can arise from a variety of optical nonlinearities, including four-wave mixing, stimulated Raman scattering, photon echoes, we will focus on comparison to the Brillouin scattering mechanism\cite{Zeldovich}. In this case one considers a system with coupling of optical fields to excitations of the matter. Examples include phonon
excitations, molecular vibrational and rotational excitations. Corresponding coupling can be represented as a scattering process
\begin{eqnarray}
V_{\rm int} = \lambda E_L E_S^* Q_B^* + {\rm c.c.}
\end{eqnarray}
where $E_L$ is the pump laser field, $E_S$ is the stimulated scattered light, and $Q_B$ describes a matter excitation amplitude, such as a phonon or a molecular vibration. Stimulated scattering leading to optical wave front reversal 
can be understood as follows \cite{Zeldovich}: after the system is pumped by laser light at $\omega_L$, it can generate
a smaller frequency photon $\omega_s$ and excitation $\Omega_B$, provided that condition $\omega_L = \omega_s+\omega_B$ is satisfied. Another way of describing this phenomenon is to observe that after the pump, the system becomes an amplifying medium for $E_s$ photons.
In the context of optical systems, front reversal phenomena correspond to optical frequencies $\omega_L$ and $\omega_s$ being much larger than the
matter excitation frequency $\omega_B$. 

Parametric instability of phonon-polaritons that we discuss in our paper does not distinguish between optical fields and matter excitations.
Due to strong linear coupling, all relevant fields are collective excitations of the medium: upper and lower polaritons. 
Parametric instability of pumped SiC insulator that we describe below
corresponds to polaritons pumped to the bottom of the upper polariton branch undergoing stimulated scattering into  a pair of polaritons at different energies, such that one
polariton goes to the lower branch and the other becomes a higher energy excitation in the upper branch
 (see Fig. \ref{fig schematic}). We do not have separation of energy scales, since all modes are in the mid-IR range
of tens of THz. One notable difference between our analysis and standard discussion
of the optical phase conjugation phenomena is that we focus on non-linear processes taking place near the material boundary rather than in the bulk. Hence in our case momentum perpendicular to the interface is not conserved and the system is less constrained by the phase matching condition. This argument also implies that we need to carefully account for 
evanescent solutions present near the boundary. This requires us to go beyond the standard Slowly Varying Envelope Approximation (SVEA) \cite{Fisher,Yariv} since at the interface between the medium and air the amplitude changes on the length scale smaller than the wavelength. This aspect of the problem has direct analogue in the scattering theory of electrons in
condensed matter physics. When describing electron states in metals, one is usually interested in single electron states with energies close to the Fermi energy. Hence, one can often
represent  electronic wavefunctions as 
\begin{eqnarray}
\psi(x) = A(x) e^{i k_F(x) x}
\label{Psi_semiclassical}
\end{eqnarray}
where $A(x)$ is a slowly varying amplitude (for simplicity we discuss the 1d case). This approximation works well when external parameters vary smoothly on the scale of $\lambda_F$. Approximation (\ref{Psi_semiclassical}) reduces Schr\"{o}edinger equation for the electron to the first order differential equations  in $x$, and as the boundary conditions one requires the continuity of $\psi(x)$ but imposes no conditions on $\frac{d\psi}{dx}$, since 
$\frac{d\psi}{dx} \approx k_F(x) \psi(x)$. The form (\ref{Psi_semiclassical}) is insufficient when one considers a problem of electron transport at the interface of two materials that have different $k_F$. The mismatch of $k_F$
gives rise to scattering of electrons \cite{Tinkham}, which one misses when imposing a single boundary condition of the continuity of $\psi(x)$. 

In our discussion of the pump and probe experiments in SiC we compute reflection coefficient of the probe beam using the full solution of Maxwell equations
and not resorting to the SVEA. In the latter approximation when looking at light propagation in a non-uniform system, one only imposes the continuity of one of the fields, for example, $E$ and assumes that continuity of $B$ is satisfied automatically, since it can be related to the spatial derivative of $E$ using the dielectric function. This is accurate only when parameters vary smoothly on the scale of a wavelength. At sharp interfaces, material properties change abruptly on the scale shorter than the wavelength, hence the full solution of Maxwell equations is necessary. One of the most surprising consequences of going beyond the SVEA is the discontinuities in the frequency dependence of the reflection coefficient (see Fig \ref{reflectivity}.).
We note however that under realistic experimental conditions two effects are expected to limit the non-analytic frequency dependence of the reflection coefficient and turn discontinuities into sharp crossovers. Firstly, the pump pulse has finite penetration depth into the material (see {\em e.g.} Ref.~\cite{Buzzi2019} for a discussion of the effect of Floquet driving with a finite penetration depth in superconductors; we emphasize that in insulators such as SiC the penetration depth is expected to be much larger).  Hence, the reflection of the probe pulse will have the character of reflection from a finite slab discussed in Appendix B.  Secondly, at frequencies close to the parametric instability reflection coefficient becomes large and non-linear terms for the probe pulse should be taken into account. They are expected to suppress the reflection coefficient, as was recently discussed for Josephson microwave parametric amplifiers by Sivak et al. \cite{PhysRevApplied.11.054060}.  This latter effect can be neglected for weak enough probe pulses.

\subsection{Organization of the paper}

This paper is organized as follows. 
In Sec.~\ref{Sec Floquet}, we introduce both linear and non-linear models of light-matter coupling in SiC. We first solve the linear model and discuss the upper and lower polariton excitations. We then consider the non-linear model and discuss propagation of the probe pulse following strong pumping resonant with the bottom of the upper polariton branch. 
In Sec.~\ref{Sec reflection}, we solve the Fresnel reflection problem of the medium with the large amplitude polariton mode. 
We obtain an analytic solution and find that the reflectivity has a discontinuous dependence on the probe light frequency. 
Furthermore, we numerically show that the parametric instability coincides with these discontinuities. 
In Sec.~\ref{Sec eigenmodes}, we study the optical properties arising from the parametric instability. 
By analyzing the temporal problem of the propagation of the probe light, we show that the parametric instability induces the probe pulse afterglow. 
We also show that for the pump field strength and the angle of incidence in the current experiments, the parametric instability should already be present. 
In Sec.~\ref{Sec Conclusions}, we discuss possible applications of the Floquet parametric instability phenomenon and future directions. 


\section{Effective model and Polaritons}\label{Sec Floquet}

\subsection{Linear theory and Polaritons}
\label{Subsec Polariton}

Polaritons in solids can be understood as collective excitations arising from hybridization of IR-active phonons with light. While this phenomenon
is standard textbook material (see {\em e.g.} Ref.~\cite{BoydBook}) we present a brief derivation here to establish notations for subsequent discussions.
We limit our discussion to the transverse optical (TO) polaritons in which electric field and polarization are perpendicular to the wavevector.

In the linear response regime the interaction between an IR active phonon mode and the electric field is described by the effective potential energy density
\begin{align}
	V(Q,E)={\Omega_{\rm TO}^2 \over 2} Q^2 - {\epsilon_0 (\epsilon_\infty -1 ) \over 2}  E^2
	-\eta {\bf Q}\cdot {\bf E},
	\label{linear hamiltonian}
\end{align}
where ${\bf Q}$ is the phonon amplitude, ${\bf E}$ is the electric field, $Q\equiv |{\bf Q}|$, $E\equiv |{\bf E}|$, $\Omega_{\rm TO}$ is the phonon frequency, $\epsilon_0$ is the permittivity of vacuum, and $\epsilon_\infty$ is the dielectric response at high frequency. 
The parameter $\eta$ characterizes the electric dipole moment of the phonon mode.  Equation (\ref{linear hamiltonian}) can be used to obtain the medium polarization density
\begin{align}
	{\bf P}&= -{\partial V \over \partial {\bf E}}\\
	&=\epsilon_0 (\epsilon_\infty -1) {\bf E} +\eta {\bf Q} , \label{EOM P polariton}
\end{align}
and the equation of motion for the phonon amplitude
\begin{align}
	\ddot{{\bf Q}}+\gamma \dot{{\bf Q}}&=-{\partial V \over \partial {\bf Q}}
	\nonumber \\
	&=-\Omega_{\rm TO}^2 {\bf Q} +\eta {\bf E},
	\label{EOM Q polariton}\end{align}
where $\gamma$ is the damping coefficient of the phonon. Collective modes can be found by combining the last two equations with the Maxwell equations
\begin{align}
	&{\bf \nabla}\times {\bf E}
	+\partial_t {\bf B}=0 \label{Maxwell E}\\
	&{1\over \mu_0} 	{\bf \nabla}\times {\bf B}
	-\partial_t \left(
		\epsilon_0 {\bf E}+{\bf P}
		\right)
	=0 \label{Maxwell B}.
\end{align}
It is convenient to combine the two Maxwell equations by taking a curl of \eqref{Maxwell E}, using \eqref{Maxwell B}, and substituting ${\bf P}$ from \eqref{EOM P polariton}
\begin{align}
	-\nabla^2 {\bf E}
	+{1\over c^2} \partial_t^2
		\left(
			\epsilon_{\infty}{\bf E} + {\eta \over \epsilon_0}{\bf Q} 
		\right)
	=0. \label{Maxwell combined polariton}
\end{align}
Equations (\ref{EOM Q polariton}) and (\ref{Maxwell combined polariton}) provide two coupled equations describing propagation of small amplitude electromagnetic waves inside the insulator. 
In a bulk material we take solutions
to be plane waves
\begin{align}
	{\bf Q} &= {\bf Q}_{\rm p} e^{i(\omega_{k} t + {\bf k}\cdot {\bf r})} + {\mathrm{c.c.}} \label{Q 0th}\\
	{\bf E} &= {\bf E}_{\rm p} e^{i(\omega_{k} t + {\bf k}\cdot {\bf r})} +{\mathrm{c.c.}}\label{E 0th},
\end{align}
where $\omega_{ k}$ is the collective mode frequency and ${\bf k}$ is its wave vector. 
We assume that ${\bf Q}_p$ is parallel to ${\bf E}_p$ and both are perpendicular to ${\bf k}$.
We find a linear set of equations for the Fourier amplitudes
\begin{eqnarray}
\left(
\begin{array}{ccc}
&k^2-{\omega_k^2 \over c^2} \epsilon_\infty & -{\eta \omega_k^2 \over c^2 \epsilon_0}\\
&-\eta &g(\omega_k) \\
\end{array} 
\right)
\left( \begin{array}{cc} {\bf E}_{\rm p}\\ {\bf Q}_{\rm p}\\ \end{array} \right)
=0,
\end{eqnarray}
where $k\equiv |{\bf k}|$ and $g(\omega) \equiv -\omega^2 -i\omega \gamma + \Omega_{\rm TO}^2$. 
To obtain eigenmodes we require the determinant to be zero and obtain 
\begin{align}
	k^2={\omega_{ k}^2 \over c^2}\left(\epsilon_\infty + {\eta^2 \over \epsilon_0 g(\omega_{ k})}\right) 
	\label{dispersion polariton}
\end{align}
Solutions of the last equation describe lower and upper polaritons shown in Fig.~\ref{fig schematic}. 
At $k=0$ and assuming $\eta \neq 0$, 
the frequency of the lower polariton is at $\omega=0$; that of the upper polariton is given by
\begin{align}
	\omega_{{\rm linear,} \, k=0}\equiv -{i\gamma\over 2} + \sqrt{-{\gamma^2 \over 4}+\Omega_{\rm TO}^2 +{\eta^2\over \epsilon_0 \epsilon_\infty} }.
	\label{omega p}
\end{align}
For the upper polariton at $k=0$ we find  a simple relation between the field and phonon amplitudes
\begin{align}
	{\bf E}_{\rm p} \equiv -{\eta \over \epsilon_0 \epsilon_\infty} {\bf Q}_{\rm p}	.
	\label{Ep Qp}
\end{align}

\subsection{Nonlinear interactions}

When the amplitude of phonon oscillations and electromagnetic fields become large, we need to include non-linear terms in the phonon-light interaction.
We include terms up to 4th order in the potential (\ref{linear hamiltonian}), consistent with the symmetry of the system. 
\begin{align}
	V=&{\Omega_{\rm TO}^2 \over 2} Q^2 -\left( {\epsilon_0 (\epsilon_\infty -1 ) \over 2} +\beta Q^2 \right) E^2 \non \\
	&-(\eta+\alpha Q^2) {\bf Q}\cdot {\bf E}.
	\label{Hamiltonian nonlinear}
\end{align}
We point out that equation (\ref{Hamiltonian nonlinear}) is valid when incident electric field is polarized in the ab plane of the 4H-SiC, as was the case in experiments by Cartella at al. \cite{Cartella12148}. The space group of 4H-SiC is P63mc (aka "Wurtzite crystal structure").  The corresponding point group is 6mm, which has 6 mirror planes
\footnote{See { http://pd.chem.ucl.ac.uk/pdnn/symm2/pg6mm.htm}}, 
all of them parallel to the c axis (i.e. they are all "vertical" mirror planes).   Analysis presented in Ref. \cite{Cartella12148} shows that the k=0 phonon participating in the formation of the polariton relevant to their experiments should have polarization in the ab plane and transform according to the two-dimensional irrep of 6mm called E2 
 \footnote{
See character table
 http://www.cryst.ehu.es/cgi-bin/rep/programs/sam/point.py?sg=183\&num=25}
. 
The square of E2 contains the trivial representation, hence, the term $Q^2$ can be part of the Hamiltonian.  On the other hand, the cube of E2 does not contain the trivial representation (to see this, one can raise the characters of the E2 row in the table to the 3rd power, and observe that this gives a set of characters that has zero overlap with the trivial representation).  In fact, one can easily check that only even powers of E2 contain the trivial representation.  We note that our analysis will also apply to the phonon that obeys the two dimensional irrep  E1 that transforms as the vectors x and y. In writing (16) we we have omitted terms of the form
$E^3Q$ and $Q^4$, as their coefficients are known to be small in SiC \cite{Cartella12148}. Note also the absence of $E^4$, which allows to preserve the standard form of the Maxwell equations. Nonlinearities are absorbed into the electric field dependence of the polarization.

 After including nonlinear terms we
find modified expressions for polarization and the equation of motion of the phonon
\begin{align}
P=& -{\partial V \over \partial {\bf E}}\\
=& \{ \epsilon_0 (\epsilon_\infty -1) + 2 \beta Q^2 \}{\bf E} +(\eta+\alpha Q^2){\bf Q},  \label{EOM P}
\\
	\ddot{{\bf Q}}+\gamma \dot{{\bf Q}}&=-{\partial V \over \partial {\bf Q}} 
	\nonumber\\
	=&(-\Omega_{\rm TO}^2+2\alpha {\bf Q}\cdot{\bf E}+ 2\beta E^2){\bf Q} +(\eta +\alpha Q^2){\bf E}.
	\label{EOM Q}	
\end{align}
Nonlinear polaritons can now be obtained by solving the phonon equation (\ref{EOM Q}) together with the Maxwell equation
\begin{align}
-\nabla^2 {\bf E}
+{1\over c^2} \partial_t^2
\left(
\epsilon_{\infty}{\bf E} + {\eta \over \epsilon_0}{\bf Q}
+{\alpha\over \epsilon_0} Q^2{\bf Q} 
+ 2{\beta \over \epsilon_0} Q^2{\bf E} 
\right)
=0. \label{Maxwell combined}
\end{align}
One important consequence of including nonlinear terms is the amplitude dependence of the polariton frequency.
 Another consequence is frequency mixing that we discuss in the next subsection. 

As the first step in analyzing the consequences of non-linearities, we neglect higher harmonics but consider renormalization of the polariton frequency by the finite excitation amplitude. From now on, when we refer to the upper polariton frequency at $k=0$ as $\omega_{\rm p}$, we imply that its renormalization by the finite drive amplitude has been included.
The ratio between ${\bf E}_{\rm p}$ and ${\bf Q}_{\rm p}$ is also modified by the non-linear terms and we denote it as $\nu$
\begin{align}
{\bf E}_{\rm p} \equiv \nu {\bf Q}_{\rm p}.
\label{Ep Qp nu}
\end{align}

In the context of pump and probe experiments we are interested in the situation where an upper polariton is excited by a strong pump pulse, and this is then followed 
by a much weaker probe pulse. In analyzing reflection of the probe pulse we will neglect its feedback on the pumped polariton. Furthermore, we will assume that on timescales of duration of the probe pulse, we can neglect the decay of the pumped polariton.  
Hence everywhere in the analysis below we assume that the pumped polariton is characterized by time-independent parameters ${\bf Q}_{\rm p}$, $\omega_{\rm p}$, and $\nu$. 
Note that the dependence of $\omega_{\rm p}$ and $\nu$ on the polariton amplitude ${\bf Q}_{\rm p}$ is always included in our numerical calculations. 

\subsection{Frequency mixing in the bulk}\label{mode mixing}

The general setting of pump and probe experiments has a strong pump pulse followed by a weaker probe. In this spirit we consider a problem of small amplitude electro-magnetic pulse propagating in the background of a large amplitude polariton oscillation. Our philosophy is to treat polariton oscillations
at $\omega_{\rm p}$ as providing periodic time modulation of the medium properties, which can be interpreted as Floquet driving. The Floquet modulation frequency is $2\omega_{\rm p}$, since nonlinear corrections to the linear equations of motion are always of the form $Q^2{\bf E}$ or $E^2{\bf Q}$.
The Floquet driving provides mixing between different frequencies and in general one needs to include an infinite number of harmonics. 
We limit our discussion to the lowest order mixing and consider small amplitude oscillations at two frequencies only, which satisfy the relation $\omega_{\rm s} + \omega_{\rm i} = 2 \omega_{\rm p}$. 
In the spirit of the general formalism of parametric resonance we will refer to the components of the frequencies $\omega_{\rm s}$ and $\omega_{\rm i}(\equiv 2\omega_{\rm p} - \omega_{\rm s})$ as the signal and the idler respectively. 
The approximation of only including mixing between two modes is justified when nonlinearities are not too strong. As a consistency check, we generalized the analysis presented in this paper to include more modes and verified that including them does not change any results qualitatively. We discuss the case of mixing between three modes in Sec.~\ref{Subsec threshold}.

We introduce three frequency components for the electric field and the phonon amplitude as
\begin{align}
&{\bf E}= {\bf E}_{\rm p} e^{-i \omega_{\rm p} t} 
+{\bf E}_{\kappa}(t)
e^{{\kappa}z}
+{\mathrm{{\mathrm{c.c.}}}}
\label{E expanded}\\
&{\bf Q}={\bf Q_{\rm p}} e^{-i \omega_{\rm p} t} 
+{\bf Q}_{\kappa}(t)
e^{{\kappa} z}+{\mathrm{c.c}} \label{Q expanded},
\end{align}
where, for concreteness, we assume a semi-infinite medium in the $z<0$ half-plane.  Here, ${\bf E}_{\rm p}$ and ${\bf Q}_{\rm p}$ are large fields, created by the strong pump pulse that drives the polariton, whereas ${\bf E}_\kappa$ and ${\bf Q}_\kappa$ are small fields, created by the weak probe pulse.  The latter fields are of the form,
\begin{align}
	&{\bf E}_{\kappa}(t) \equiv {\bf E}_{1} e^{-i \omega_{\rm s} t} + {\bf E}_{2}^* e^{i \omega_{\rm i} t}\\
	&{\bf Q}_{\kappa}(t) \equiv {\bf Q}_{1} e^{-i \omega_{\rm s} t} + {\bf Q}_{2}^* e^{i \omega_{\rm i}t },
\end{align}
so that ${\bf E}_1$ and ${\bf E}_2$ should be interpreted as amplitudes of the signal and idler components, respectively. 
Our goal is to solve Maxwell's equations linearized with respect to the small $\omega_{\rm s}$ and $\omega_{\rm i}$ components. This allows us to neglect their feedback on the strong polariton component at $\omega_{\rm p}$. In writing equations (\ref{E expanded}) and (\ref{Q expanded}) we changed notation for the spatial dependence of the fields to the evanescent wave form $e^{{\kappa}z}$, where ${\kappa} \equiv i k$. This is done for convenience of subsequent discussions. Note that the frequency components $\omega_{\rm s}$ and $\omega_{\rm i}$ have the same spatial structure because they correspond to different Fourier components of the same Floquet eigenmode and the pumped polariton is at $k=0$.
Having the same spatial structure for different frequency components should be contrasted to solutions of the Maxwell equations in stationary media, 
where each frequency $\omega$ has its own wave vector $\kappa(\omega)$. 
In the Floquet formalism a single eigenmode is constructed as a superposition of several frequency components, all sharing the same wavevector. 

After straightforward calculations, which are presented in the Appendix A, we obtain linear equations for the imaginary wavevectors $\kappa$ of Floquet eigenmodes.  
\begin{eqnarray}
\left(
\begin{array}{ccc}
&-\kappa^2+\tilde{\kappa}_{\rm s}^2 & -\Gamma_1\\
&-\Gamma_2 &-\kappa^2+\tilde{\kappa}_{\rm i}^2  \\
\end{array} 
\right)
\left( \begin{array}{cc} {\bf E}_{1}\\ {\bf E}_{2}^*\\ \end{array} \right)
=0,
\label{Maxwell matrix}
\end{eqnarray}
where 
\begin{align}
\tilde{\kappa}_{\rm s}  \equiv&  \sqrt{-{\omega_{\rm s}^2\over c^2} \tilde{\epsilon}(\omega_{\rm s})}\,,
\quad\tilde{\kappa}_{\rm i} \equiv  \sqrt{-{\omega_{\rm i}^2\over c^2} \tilde{\epsilon}(-\omega_{\rm i})}\,, \label{tildeks}\\	
\Gamma_1\equiv& {\omega_{\rm s}^2 \over c^2} Q_{\rm p}^2 G(\omega_{\rm s})\,,\quad
\Gamma_2\equiv {\omega_{\rm i}^2 \over c^2} Q_{\rm p}^2 G(\omega_{\rm s})\,,\\
\tilde{\epsilon}(\omega)
\equiv& \epsilon_\infty + {\eta^2 \over \epsilon_0 \tilde{g}(\omega)}  + F(\omega) |Q_{\rm p}|^2 \,, 
\label{tilde epsilon} \\
\tilde{g}(\omega) \equiv& g(\omega)+(-4\beta\nu^2 + 12 \alpha \nu)|Q_{\rm p}|^2\,,\\
F(\omega) \equiv& {1\over \epsilon_0}\left[
4\beta + {4\eta\over \tilde{g}(\omega)}(3\alpha-4\beta \nu)
\right]\,,\\
G(\omega) 
\equiv &
{1 \over \epsilon_0}\Bigg[
2\beta+\left({\eta \over \tilde{g}(\omega)}+{\eta \over \tilde{g}(\omega-2\omega_{\rm p})}\right)(3\alpha -4 \beta \nu ) \non \\
&+{2\eta^2 \over \tilde{g}(\omega) \tilde{g}(\omega-2\omega_{\rm p})} (-3\alpha \nu + \beta \nu^2)
\Bigg]\,.\label{Gdef}
\end{align}

\begin{figure*}
		\includegraphics[width=2\columnwidth]{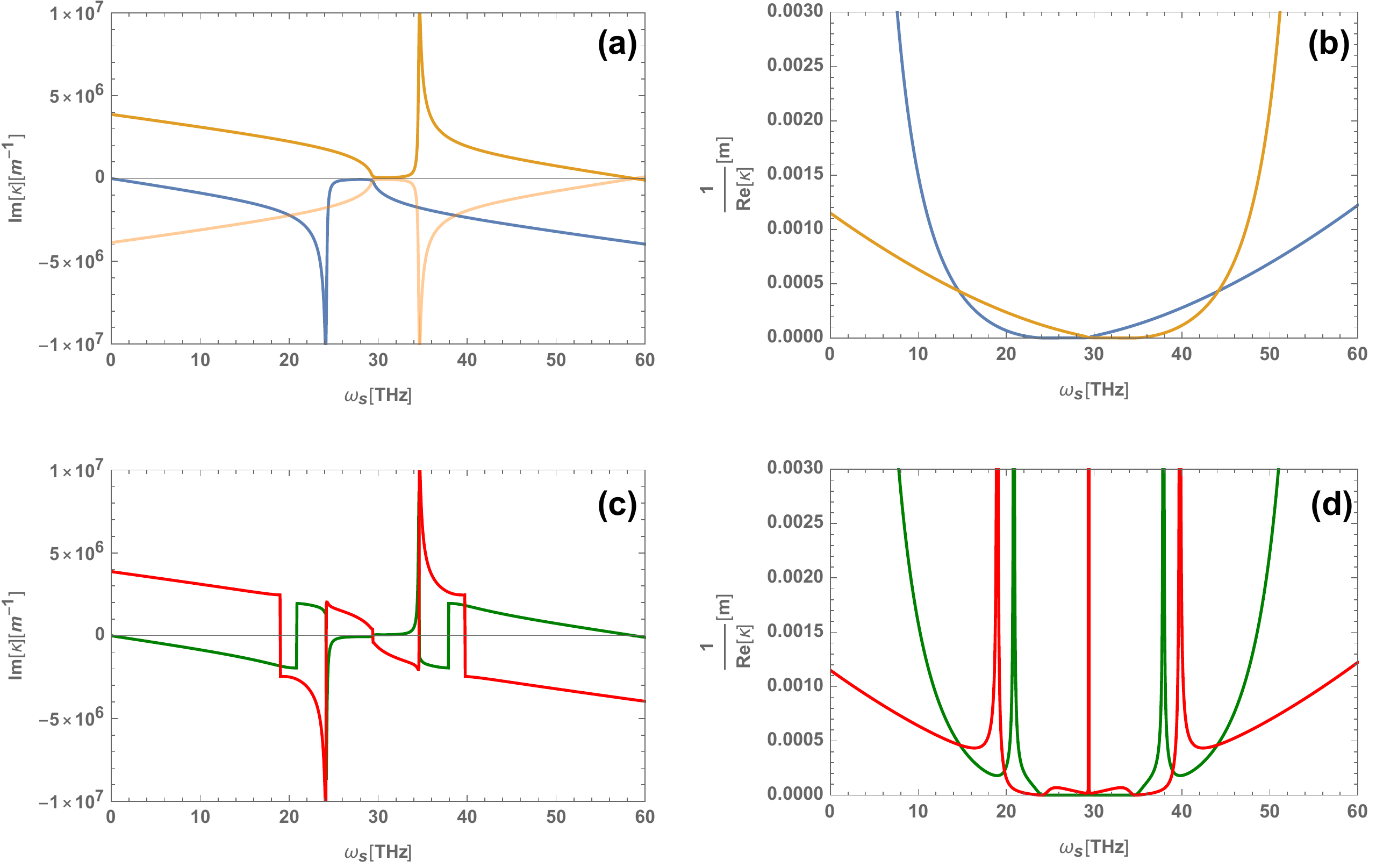}
				\caption{(a) Wave numbers without the Floquet frequency mixing (vertical) {\em vs} frequency $\omega_{\rm s}$ (horizontal). $\img[\tilde{\kappa}_{\rm s}]$ (blue) and $\img[\tilde{\kappa}_{\rm i}]$ (dark yellow) are shown. $\Omega_{\rm TO}=24$THz and $\omega_{p}=29.4$THz. They give the dispersion relation of the polariton. In order to see the phase-matching condition \eqref{Kappas_Kappai_condition}, $-\img[\tilde{\kappa}_{\rm i}]$ (light yellow) is also shown. At $\omega_{\rm s}=20,24,30,35$ and 39THz, the phase-matching condition is satisfied. (b) Localization lengths without the hybridization(vertical) {\em vs} frequency (horizontal). $1/\real[\tilde{\kappa}_{\rm s}]$ (blue) and $1/\real[\tilde{\kappa}_{\rm i}]$ (yellow) are shown. Since $\img[\tilde{\kappa}_{\rm s}]$ has a gap between the upper and lower branches at $\Omega_{\rm TO}<\omega_{\rm s}<\omega_{\rm p}$, there are no propagating modes in this region but only evanescent waves. Thus, the localization length becomes almost zero. $1/\real[\tilde{\kappa}_{\rm i}]$ also becomes short at $\omega_{p}<\omega_{\rm s}<2\omega_{\rm p}-\Omega_{\rm TO}$. 
				(c) Analysis of imaginary wave numbers $\kappa_\pm$ after including Floquet frequency hybridization. $\img[\kappa_+]$ (green) and $\img[\kappa_-]$ (red) are shown. At the frequencies where \eqref{Kappas_Kappai_condition} is satisfied, {\em i.e.}, $\omega_{\rm s}=20,24,30,35$ and 39THz, strong mixing of the signal and idler light occurs. (d) Localization lengths of Floquet eigenmodes as a function of frequency $\omega_{\rm s}$(horizontal). Both $\real[\kappa_+]$ (green) and $\real[\kappa_-]$ (red) are presented. Strong Floquet mixing of the signal and idler frequencies leads to localization length becoming infinite at three frequencies: $\omega_{\rm s}=$20, 29 and 39THz. At these frequencies light propagates inside the Floquet medium without decay. Dissipation is compensated by pumping. At $\omega_{\rm s}$=24THz and 35THz, the localization length becomes zero. This corresponds to polaritons localized at the surface of the insulator.}
		\label{fig kappas}
\end{figure*}

Equations (\ref{tildeks})-(\ref{Gdef}) have been arranged to provide a simpler physical interpretation.
The function $\tilde{\epsilon}(\omega)$ has been defined as an effective dielectric constant when
solving the linearized phonon-Maxwell equations in the presence of strong oscillations at $\omega_{\rm p}$
but without including the frequency mixing between the signal and idler Fourier components. This dielectric
constant function is then used to define the ``imaginary wave vectors"  $\tilde{\kappa}_{\rm s}$ and  $\tilde{\kappa}_{\rm i}$ corresponding to $\omega_{\rm s}$ and $\omega_{\rm i}$. 
Figure \ref{fig kappas}a shows the dependence of $\img[\tilde{\kappa}_{\rm s}]$ and $\img[\tilde{\kappa}_{\rm i}]$ on $\omega_{\rm s}$ (in the latter case $\omega_{\rm i} = 2 \omega_{\rm p}-\omega_{\rm s}$). Material parameters used in our analysis have
been taken from Ref.~\cite{Cartella12148} and are summarized in Table~\ref{Table parameters}. The function $\img[\tilde{\kappa}_{\rm s}(\omega_{\rm s})]$ is qualitatively similar to the dispersion for linear polaritons. We can identify the lower and upper polariton branches below and above the bare phonon frequency $\Omega_{\rm TO}$. 

From the solution of (\ref{Maxwell matrix}) we find the Floquet eigenmodes 
\begin{align}
\kappa_{\pm}^2 
\equiv
{\tilde{\kappa}_{\rm s}^2 + \tilde{\kappa}_{\rm i}^2 \over 2}
\pm \sqrt{
	\left({\tilde{\kappa}_{\rm s}^2-\tilde{\kappa}_{\rm i}^2 \over 2}\right)^2 + \Gamma_1 \Gamma_2
},
\label{kappa pm}
\end{align}
The eigensolutions of $({\bf E}_1,{\bf E}_2^*)$ satisfy
\begin{align}
	{\bf E}_{2\pm}^* = 
	\alpha_\pm {\bf E}_{1\pm} \label{relation Es Ei}
\end{align}
where $({\bf E}_{1+},{\bf E}^*_{2+})$ and $({\bf E}_{1-},{\bf E}^*_{2-})$ are the eigensolutions of $\kappa_+$ and $\kappa_-$, respectively, and 
\begin{align}
	\alpha_\pm \equiv {-\kappa_\pm^2 + \tilde{\kappa}_{\rm s}^2 \over \Gamma_1}.
\end{align}
To summarize the discussion in this section, the Floquet eigensolutions are expressed as superpositions of two waves, of frequencies $\omega_{\rm s}$ and $\omega_{\rm i}$: 
\begin{align}
	&{\bf E}_\pm(t) \equiv {\bf E}_{1\pm} (e^{-i \omega_{\rm s} t} + \alpha_\pm e^{i \omega_{\rm i} t}) \label{E floquet}
\end{align}

Results of numerical calculations of $\kappa_{\pm}$ are shown in Figs.~\ref{fig kappas}c and d.  Note that to compute the renormalization of $\omega_{\rm p}$ due to the finite amplitude of the
pumped polariton, we need to solve $\real[\tilde{\epsilon}(\omega_{\rm p})]=0$. For example, we find $\omega_{\rm p}=$ 29.4THz when $|{\bf E}_{\rm p}|$ = 8MV/cm. This should be compared to the
bare polariton frequency of 30.5THz

\begin{table}[htb]
	\begin{tabular}{|l|c l|} \hline
		$\Omega_{\rm TO}$ & $24.0$ &$[{\rm THz}]$\\ \hline
		$\gamma$ & $0.2$ &$[{\rm THz}]$\\ \hline
		$\epsilon_\infty$ & $5.91$& \\ \hline
		$E_{\rm p}$ & $8.00 \times 10^2$ & [MV/m] \\ \hline
		$\eta$ & $8.39\times 10^{8}$ &$[{\rm As}/{\rm kg}^{1\over2}m^{3 \over 2}]$\\ \hline
		$\alpha$ & $7.11\times 10^{25}$ &$[{\rm As /kg}^{3\over2}m^{1 \over2}]$\\ \hline
		$\beta$ & $1.69\times 10^{7}$ &$[{\rm A}^2 {\rm s}^4 /{\rm kg}^{2}m^2]$\\ \hline
		$\nu$ & $9.82 \times 10^{11}$& [MV/${\rm kg}^{1\over 2}m^{1\over 2}$] \\ \hline
	\end{tabular}
	\caption{Parameters used in our numerical calculation. They are taken from Ref.~\cite{Cartella12148}, in which terahertz optical amplification in SiC is experimentally observed.}\label{Table parameters}
\end{table}

\subsection{Discussion of Floquet eigenmodes}
\label{KappaDiscontinuitySection}

\subsubsection{Discontinuities and phase matching condition}

\begin{figure}[htp]
\begin{center}
		\includegraphics[width=0.4\textwidth]{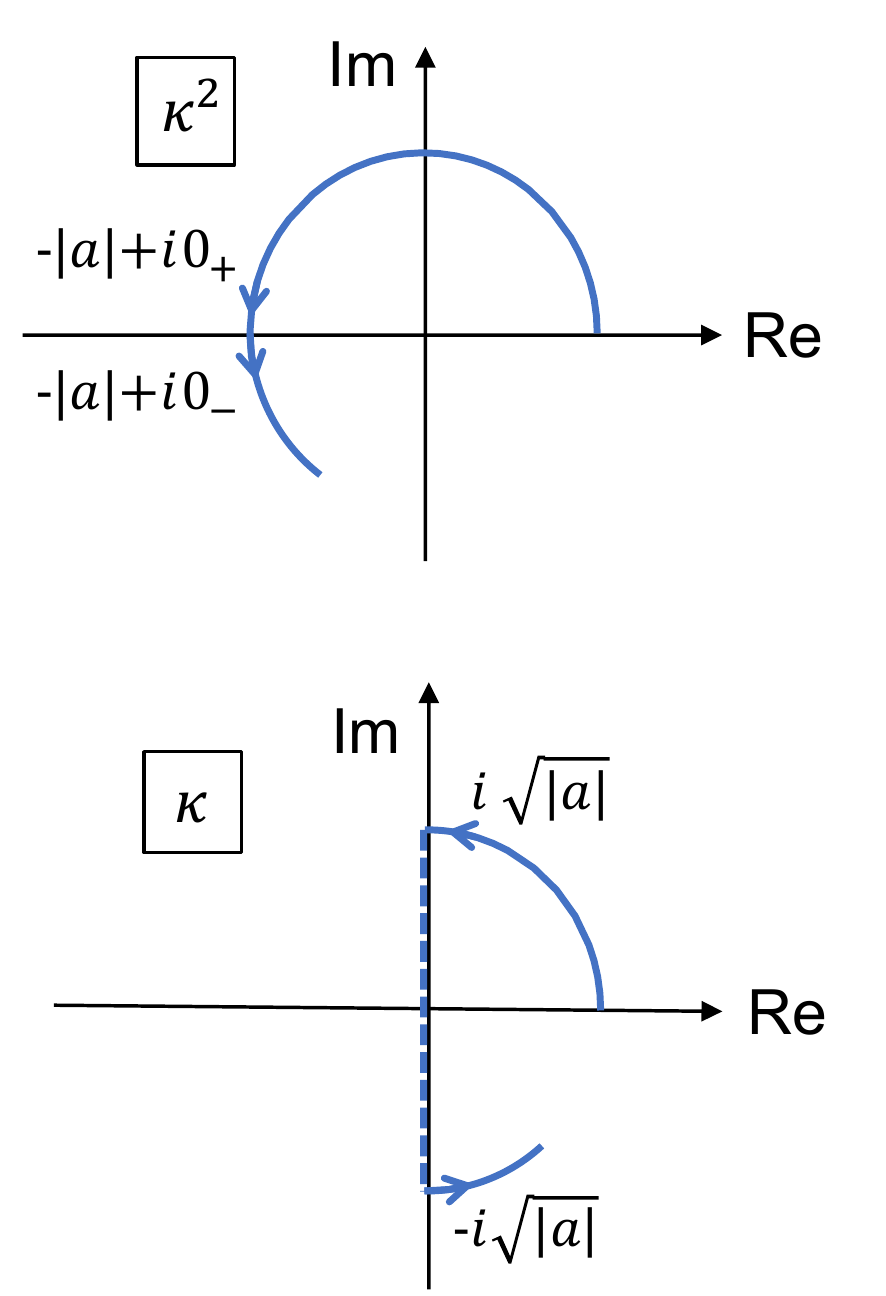} 
	\caption{Discontinuity of $\img \, \kappa_{\pm}$ can be related to switching the branch in the square root 
		when computing $\kappa_{\pm}$ from equation \eqref{kappa pm}. This equations specifies $\kappa_{\pm}^2$ and the correct choice
		of the branch is determined by the condition $\real \, \kappa_{\pm} \geq 0$.
 }
	\label{kappajump}
\end{center}
\end{figure}

The most surprising feature of the computed eigenmodes, shown in Fig.~\ref{fig kappas}c, is the 
discontinuous dependence of $\img[\kappa_\alpha]$ at the frequencies 20, 30, and 39 THz (here $\alpha$ is either $+$ or $-$).  From a mathematical point of view, these discontinuities arise because Eq.~(\ref{kappa pm}) is an equation for 
$\kappa_{\alpha}^2$, and  $\kappa_{\alpha}$ itself has an overall sign ambiguity.  For a semi-infinite sample, we fix the sign using the condition
$\real [\kappa_{\alpha}]\geq 0$, so that in the light reflection problem we can construct solutions decaying from the interface into the bulk of the material. In Fig.~\ref{kappajump} we show that when $\kappa_{\alpha}^2$ crosses the negative real axis,  $\img [\kappa_{\alpha}]$ jumps discontinuously, whereas $\real [\kappa_{\alpha}]$ has a kink as it touches zero while always remaining nonnegative.  Figure \ref{fig kappas}d shows $\frac{1}{\real[\tilde{\kappa}_\alpha]}$, which can be understood as the localization length of the Floquet eigenmodes. In agreement with our discussion, the points where $\img[\tilde{\kappa}_\alpha]$ is discontinuous coincide with divergences in $1/{\real[\tilde{\kappa}_\alpha]}$, implying that the Floquet solutions at these frequencies are perfectly traveling waves with no decay.  As we will show in the following, for a semi-infinite medium, the discontinuities in $\img[\tilde{\kappa}_\alpha]$ result in discontinuous jumps in the reflection coefficient $R_s(\omega)$.

For a finite slab geometry, one may object that fixing the sign of  $\kappa_{\alpha}$ is not physical, since solutions with both signs (corresponding both to exponential decay and growth as a function of $z$) must be kept in the reflection problem.  However, as the slab becomes thicker, the coefficient of the growing solution becomes exponentially small and, for a thick enough slab, it plays a negligible role in the reflection problem.  In other words, we expect that for a thick sample, the discontinuities in $R_s(\omega)$ are replaced by rapid crossovers near the frequencies of the discontinuities in $\img[\kappa_\alpha]$.  The crossover regions are given the frequency windows over which the localization lengths in Fig.~\ref{fig kappas}d exceed the sample thickness.  The widths of these windows decay as the inverse slab thickness.  We verified that this is indeed the case through explicit computation of reflection from finite slab geometries.

Let us now discuss the physical conditions required for the discontinuities in $\img[\kappa_\alpha]$ to occur. Comparison of Figs.~\ref{fig kappas}a and c shows that these occur near frequencies where 
\begin{eqnarray}
\img [\tilde{\kappa}_{\rm s}] = - \img [\tilde{\kappa}_{\rm i}].
\label{Kappas_Kappai_condition}
\end{eqnarray} 
Comparing with equation (\ref{kappa pm}), we observe that when condition (\ref{Kappas_Kappai_condition}) is satisfied, Floquet frequency mixing 
set by the coefficients $\Gamma_1$ and $\Gamma_2$ becomes strong and significantly shifts the mode vectors $\kappa_{\pm}$, which become very different from $\kappa_{\{s,i\}}$.  Condition (\ref{Kappas_Kappai_condition}) can be understood as the phase matching
condition: the two frequency components have the same spatial structure and can be efficiently mixed by the spatially uniform
but time varying modulation of system parameters at frequency $\omega_{\rm p}$. We also observe that the discontinuities occur in the frequency regime where $| \img [\tilde{\kappa}_{\{s,i\}}] | \gg | \real [\tilde{\kappa}_{\{s,i\}}] |$, so that the signal and idler modes are primarily traveling waves with relatively weak damping.

In addition to the discontinuities discussed above, note that $\img[\kappa_\alpha]$ in Fig.~\ref{fig kappas}c also displays sharp crossovers accompanied by peaks at the frequencies 24 and 35 THz.  The crossover at 24 THz occurs at the original phonon mode frequency $\Omega_{\rm TO}$, whereas that at 35 THz is an ``idler satellite" peak at frequency $2\omega_{\rm p} - \Omega_{\rm TO}$.  These crossovers occur because $1\over\tilde{g}(\omega)$ has a  sharp maximum at $\Omega_{\rm TO}$.  The phonon dissipation $\gamma$ controls the sharpness of these crossovers, as well as their width.

\subsubsection{Eigenmodes away from frequency matching condition}

Far from the special frequencies satisfying the phase matching condition, we have
\begin{align}
\sqrt{|\Gamma_1\Gamma_2|} \ll  |\tilde{\kappa}_{\rm s}^2 - \tilde{\kappa}_{\rm i}^2|	\label{phase matching precise}
\end{align}
and hybridization between the signal, $\omega_{\rm s}$, and idler, $\omega_{\rm i}$, modes is weak.
We find that for $|\tilde{\kappa}_{\rm s}|<|\tilde{\kappa}_{\rm i}|$ 
\begin{align}
\kappa_+ \simeq \tilde{\kappa}_{\rm s} \label{weak hyb 1}\\
\kappa_- \simeq \tilde{\kappa}_{\rm i}\\
|\alpha_+| \ll 1\\
|\alpha_-| \gg 1	 \label{weak hyb 4}
\end{align}
whereas for $|\tilde{\kappa}_{\rm s}|>|\tilde{\kappa}_{\rm i}|$ the roles of $+$ and $-$ are reversed.


\section{Fresnel-Floquet reflection problem} \label{Sec reflection}

\subsection{Solution of Fresnel-Floquet problem}

\begin{figure}[htp]
	\begin{center}
		\includegraphics[width=0.4\textwidth]{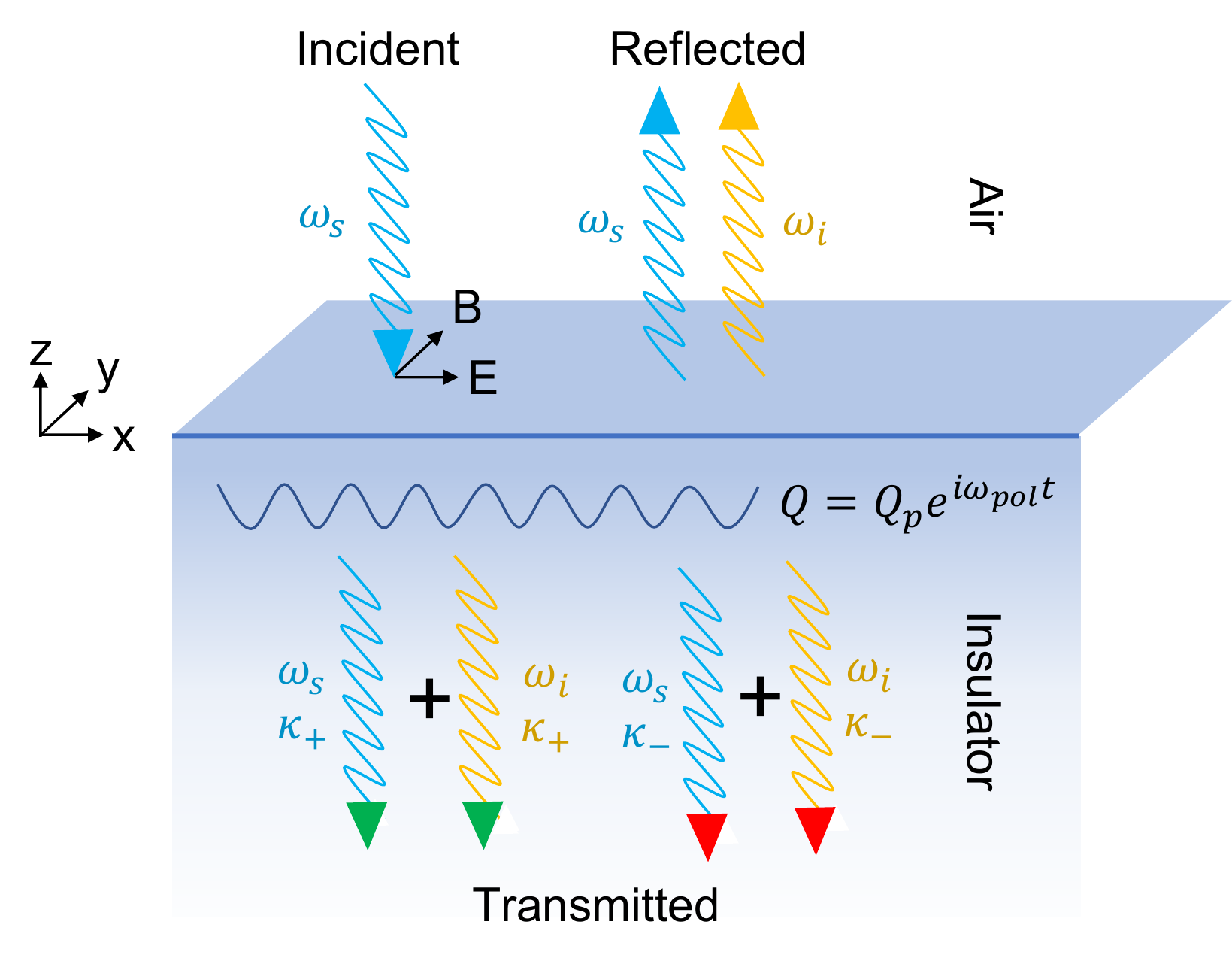}
		\caption{Schematic figure of the Fresnel-Floquet light reflection. The incident light has a frequency $\omega_{\rm s}$ while the reflected light has components of frequencies $\omega_{\rm s}$ and $\omega_{\rm i}$ due to the parametric amplification process. Inside the slab, large amplitude of phonon mixes them; the Floquet eigenmodes are the superpositions of these two components which take the same spatial dependence. }
		\label{fig_fresnel_floquet}
	\end{center}
\end{figure}

We now consider the problem of light reflection from an insulator with an excited polariton mode. Here we present the analysis for a semi-infinite slab of material and normal angle of incidence. Generalizations to cases of a slab of finite thickness and arbitrary angle of incidence are discussed in Appendix~\ref{Subsec nonnormal}. The main difference between our system and the canonical Fresnel reflection problem is frequency mixing. An incident beam at frequency $\omega_{\rm s}$ excites an electromagnetic response in the insulator at two frequencies: $\omega_{\rm s}$ and $\omega_{\rm i} = 2 \omega_{\rm p} - \omega_{\rm s}$. Therefore for both the transmitted and reflected beams we need to include both frequency components (see Fig. \ref{fig_fresnel_floquet}). When describing light penetrating into the nonequilibrium insulator we need to use eigenmodes of the system. Hence we will decompose the transmitted wave into Floquet eigenmodes $\kappa_+$ and  $\kappa_-$.  We then have the following decomposition of the fields: 

In the air, $z>0$, for linearly polarized light,
\begin{eqnarray}
\left( \begin{array}{c} {\bf E}(z,t) \\ {\bf B}(z,t) \end{array} \right)
&=& \left( \begin{array}{c}  {E}_{\rm inc}(z,t) \hat{e}_x \\ {B}_{\rm inc}(z,t) \hat{e}_y \end{array} \right)
+
 \left( \begin{array}{c}  {E}_{\rm ref}(z,t) \hat{e}_x \\ {B}_{\rm ref}(z,t) \hat{e}_y \end{array} \right)\,,
 \nonumber
\end{eqnarray}
where
\begin{eqnarray}
 \left( \begin{array}{c}  {E}_{\rm inc}(z,t) \\ {B}_{\rm inc}(z,t)  \end{array} \right)
& =& E_0\left( \begin{array}{c}  1 \\ \sqrt{\epsilon_0 \mu_0} \end{array} \right)\, e^{ -i \omega_{\rm s} t} e^{\kappa_{\rm s} z}\,,
\label{reflection incident} \\
 \left( \begin{array}{c}  {E}_{\rm ref}(z,t) \\ {B}_{\rm ref}(z,t)  \end{array} \right)
 &=& r_{\rm s}E_0 \, \left( \begin{array}{c}  1 \\ -\sqrt{\epsilon_0 \mu_0} \end{array} \right)\, e^{ -i \omega_{\rm s} t} e^{ -\kappa_{\rm s} z}\
 \nonumber\\
&+ & r_{\rm i} E_0\, \left( \begin{array}{c}  1 \\ -\sqrt{\epsilon_0 \mu_0} \end{array} \right)\, e^{ i \omega_{\rm i} t} e^{-\kappa_{\rm i} z} \label{EB ref}\,.
\end{eqnarray}
Here, 
\begin{align}
	&\kappa_{\rm s}\equiv -i{\omega_{\rm s}\over c}  
	\label{ks_vac}\\	
	&{\kappa}_{\rm i}\equiv i{\omega_{\rm i}\over c}\label{ki vac}.
\end{align}
Note that we take the opposite signs for $\kappa_{\rm s}$ and $\kappa_{\rm i}$ in equations  (\ref{ks_vac}) and (\ref{ki vac}) so that both describe waves traveling upward {\em i.e.} reflected from the surface of the insulator.   The complex coefficient $r_{\mathrm s}$ appearing in Eq. (\ref{EB ref}) is the reflection coefficient for the signal, whereas $r_{\mathrm i}$ describes the amplitude of the reflected idler, both of which are determined below.

In the insulator, for $z<0$,
\begin{eqnarray}
\left( \begin{array}{c}  {\bf E}(z,t) \\ {\bf B}(z,t)  \end{array} \right) 
&=& t_+ \, \left( \begin{array}{c}  {E}_{+}(t)\hat{e}_x  \\ {B}_{+}(t)\hat{e}_y \end{array} \right)
e^{ \kappa_+ z}
\nonumber\\
&+&
t_- \, \left( \begin{array}{c}  {E}_{-}(t)\hat{e}_x  \\ {B}_{-}(t) \hat{e}_y \end{array} \right)
e^{ \kappa_- z} \label{EB trans}
\end{eqnarray}
Expressions for $E_{\pm}(t)$ should be taken from \eqref{E floquet}. By substituting $E_{\pm}(t)$ into \eqref{Maxwell E} we find that the magnetic field in \eqref{EB trans} is given by 
\begin{align}
	B_{\pm}(t)\equiv i\kappa_\pm E_{1\pm} \left({e^{-i \omega_{\rm s} t} \over \omega_{\rm s}}  -\alpha_\pm {e^{i \omega_{\rm i} t}  \over \omega_{\rm i}}  \right).
\end{align}
The coefficients $r_{\rm{s/i}}$, $t_{\pm}$ can be determined using the boundary conditions for the electric and magnetic fields. 
We require continuity of the electric and magnetic fields at the boundary (note that, for normal incidence, the fields are parallel to the interface), we find
\begin{eqnarray}
 (e^{- i \omega_{\rm s} t} &+& r_{\rm s} e^{- i \omega_{\rm s} t}+ r_{\rm i} e^{ i \omega_{\rm i} t} ) E_0 =
 t_+ E_+(t) + t_- E_-(t) 
 \nonumber\\
\sqrt{\epsilon_0 \mu_0} &(& e^{- i \omega_{\rm s} t} - r_{\rm s} e^{- i \omega_{\rm s} t} - r_{\rm i} e^{ i \omega_{\rm i} t}) E_0 
\nonumber\\
&=&t_+ B_+(t) + t_- B_-(t)
 \label{reflection boundary}
\end{eqnarray}

Solving \eqref{reflection incident}-\eqref{reflection boundary}, we obtain the reflection and transmission coefficients. 
\begin{align}
	r_{\rm s} 
	= {
	\alpha_-(\kappa_{\rm i}+\kappa_-)(\kappa_{\rm s}-\kappa_+)-\alpha_+(\kappa_{\rm i}+\kappa_+)(\kappa_{\rm s}-\kappa_-)
	\over
	\alpha_-(\kappa_{\rm i}+\kappa_-)(\kappa_{\rm s}+\kappa_+)-\alpha_+(\kappa_{\rm i}+\kappa_+)(\kappa_{\rm s}+\kappa_-)	
	} \label{rs} \\
	r_{\rm i}
	= {
	2\alpha_-\alpha_+ \kappa_{\rm s}(-\kappa_++\kappa_-)
	\over
	\alpha_-(\kappa_{\rm i}+\kappa_-)(\kappa_{\rm s}+\kappa_+)-\alpha_+(\kappa_{\rm i}+\kappa_+)(\kappa_{\rm s}+\kappa_-)	
	}\label{r_i}
	\\
	t_{+}
	= {
	2\alpha_- \kappa_{\rm s}(\kappa_{\rm i}+\kappa_-)
	\over
	\alpha_-(\kappa_{\rm i}+\kappa_-)(\kappa_{\rm s}+\kappa_+)-\alpha_+(\kappa_{\rm i}+\kappa_+)(\kappa_{\rm s}+\kappa_-)	
	}\\
	t_{-}
	= {
	-2\alpha_+ \kappa_{\rm s}(\kappa_{\rm i}+\kappa_+)
	\over
	\alpha_-(\kappa_{\rm i}+\kappa_-)(\kappa_{\rm s}+\kappa_+)-\alpha_+(\kappa_{\rm i}+\kappa_+)(\kappa_{\rm s}+\kappa_-)	
	}
	\label{t-}
\end{align}

\begin{figure}[h]
	\begin{center}
           \includegraphics[width=0.9\columnwidth]{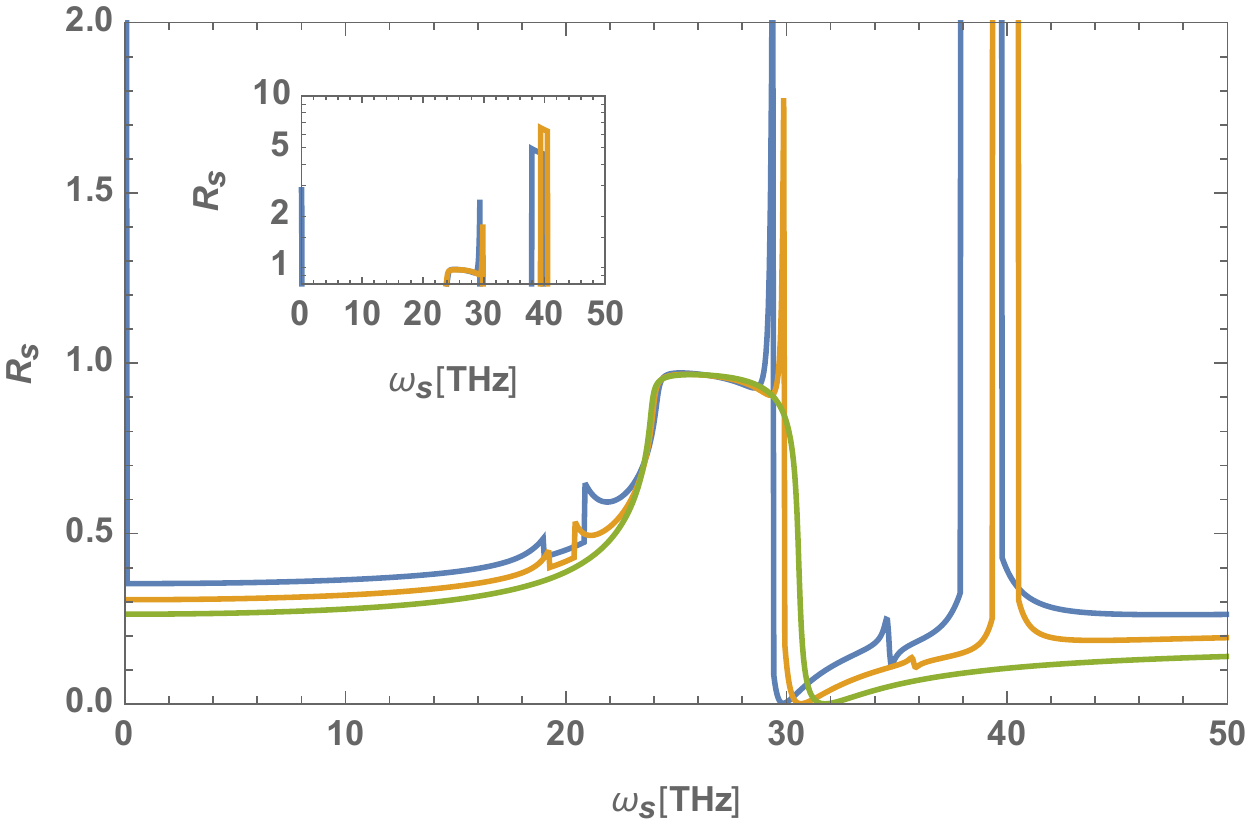}
    \caption{Reflectivity $R_{\rm s}\equiv|r_{\rm s}|^2$ (vertical) vs frequency $\omega_{\rm s}$ of the incident light (horizontal) obtained using Eq.~\eqref{rs}. Different lines correspond to different pump field amplitudes; $|{\bf E}_0|=0$ (green), 5MV/cm(orange), and 8MV/cm (blue).
    We see discontinuities and peaks of the reflectivity when $|{\bf E}_0|\neq$ 0. 
    The discontinuities can be seen around three frequencies, $\omega_{\rm s}=20,29,$ and 39THz. 
    They indicate the presence of the parametric instability discussed in Sec.~\ref{Subsec discontinuity}. 
    There is a strong peak at $\omega_{\rm s}=35$THz and a much weaker one at $\omega_{\rm s}=24$THz (not visible on this scale). They correspond to the surface modes discussed in Sec.~\ref{Subsec SPhP}. 
    (Inset) Logarithmic plot of $R_{\rm s}$ outside the range of the linear plot. Around 29 and 39 THz, $R_{\rm s}$ takes large values. 
    } 
    \label{reflectivity}
    \end{center}
\end{figure}

\subsection{Numerical results for the reflectivity $r_{\rm s}$}

Results of numerical calculations of the reflectivity
\begin{align}
	R_{\rm s} \equiv |r_{\rm s}|^2
\end{align}
for three values of the
pump field amplitude, $|{\bf E}_{\rm p}|$=0, 5, and 8 MV/cm are presented in Fig.~\ref{reflectivity}. Due to non-linearities the
frequency of the polariton mode differs somewhat in the three cases, taking values 
$\omega_{\rm p}=$30.5, 29.9, and 29.4THz respectively. 
System parameters are taken from Table \ref{Table parameters}.
The green line corresponds to the situation without the pump field, {\em i.e.} $|{\bf E}_{\rm p}|=0$. 
Note that in this case the incident light is reflected almost perfectly in the frequency gap between the two branches of polaritons. This spectral region $\Omega_{\rm TO}<\omega_{\rm s}<\omega_{\rm p}$ is called the reststrahlen band. 
Orange and blue lines correspond to $R_{\rm s}$ for $|{\bf E}_{\rm p}|=5$MV/cm and 8MV/cm, respectively. 
For finite $|E_{\rm p}|$ the reststrahlen band is still visible, although its width is reduced due to renormalization of $\omega_{\rm p}$ by non-linear effects. 
At finite pump fields $R_{\rm s}$ starts to show non-analytic dependence on frequency. 
At frequencies $\omega_{\rm s}=20,29,39$ THz, we observe abrupt jumps in the reflection coefficient. 
At $\omega_{\rm s}$= 35THz, we observe a peak in the reflection coefficient but no 
discontinuity (see Figs.~\ref{fig Rinf} and \ref{fig surfacemode} for a more detailed information about the frequency dependence of $R_{\rm s}(\omega)$ around these points). 
The jumps and the peak of $R_{\rm s}$ at these frequencies are obviously related to the discontinuities and the crossover in $\kappa_\pm$, which in turn
originates from the fact that phase matching conditions are satisfied at these frequencies. 
We point out that for different pump intensities singularities at 35 and 39THz occur at slightly different frequencies because $\omega_{\rm i}$ depends on the value of $\omega_{\rm p}$. 

Far from the special frequencies that satisfy the phase matching condition 
\eqref{Kappas_Kappai_condition},
$r_{\rm s}$ is essentially the same as in the linear model regardless of the value of $|{\bf E}_{\rm p}|$.
In this case, we can use \eqref{weak hyb 1} and  approximate \eqref{rs} as 
\begin{align}
r_{\rm s} 
\simeq {
	\kappa_{\rm s}-\tilde{\kappa}_{\rm s}	\over
	\kappa_{\rm s}+\tilde{\kappa}_{\rm s}	}. \label{rs reduced}
\end{align}
This is the familiar Fresnel's formula. 
Since $\img[\kappa_{\rm s}]$ and $\img[\tilde{\kappa}_{\rm s}]$ have the same sign, 
and $\real[\kappa_{\rm s}]=0$, 
equation \eqref{rs reduced} shows that in this case $R_{\rm s}$ is continuous and smaller than one. 

We note that the value of the pump electric field used in experiments by Cartella et al. was as high as 8.68MV/cm. Hence we expect the Floquet parametric instability
to be easily achievable in experiments with SiC insulator.

\subsection{Understanding $r_{\rm s}$: non-analytic behavior and lasing modes}\label{Subsec discontinuity}

The most surprising feature of the results presented in Fig.~\ref{reflectivity} are the discontinuities in the frequency dependence of the reflectivity $R_{\rm s}$. 
There are two close pairs of discontinuities, with the first pair at 19/20.5 THz and the second pair at 38/39.5THz. There is also an isolated spike at frequency 29THz. These discontinuities are a
direct consequence of the discontinuous dependence of $\kappa_\pm$ on $\omega$ discussed in Section \ref{KappaDiscontinuitySection}. Previously we related such discontinuities to the existence of special frequencies that satisfy phase matching conditions for the signal and  idler
modes, at which Floquet modulation gives rise to strong mixing between the modes. We will now discuss that behind these discontinuities lies an even more dramatic physical effects: parametric instabilities of the Floquet medium. In the case of a close pair of discontinuities, {\em e.g.} the 38/39.5THz pair, unstable modes appear in the whole range of frequencies between the two. 
In the case of a single spike, parametric instability occurs at the same frequency. As we discuss below these modes are spatially extended inside the slab. 

We now review the arguments which allow to identify parametric instability as the origin of singular behavior of $R_{\rm s}$. Let us consider reflectivity from the perspective of a response function of a physical system. It is useful to extend the definition of reflection coefficient from real to complex frequencies. We can look for poles of $R_{\mathrm{s/i}}(\omega)$ and identify them with collective modes of the system. In a static dissipative medium such poles should have negative imaginary part of the frequency indicating finite decay rate. For gain media one can find poles with positive imaginary part which correspond to parametric instability instabilities\cite{Dorofeenko2012}. The relation of these poles and the lasing phenomena are presented in Sec.~\ref{Sec eigenmodes}. 

\begin{figure*}
	\includegraphics[width=2 \columnwidth]{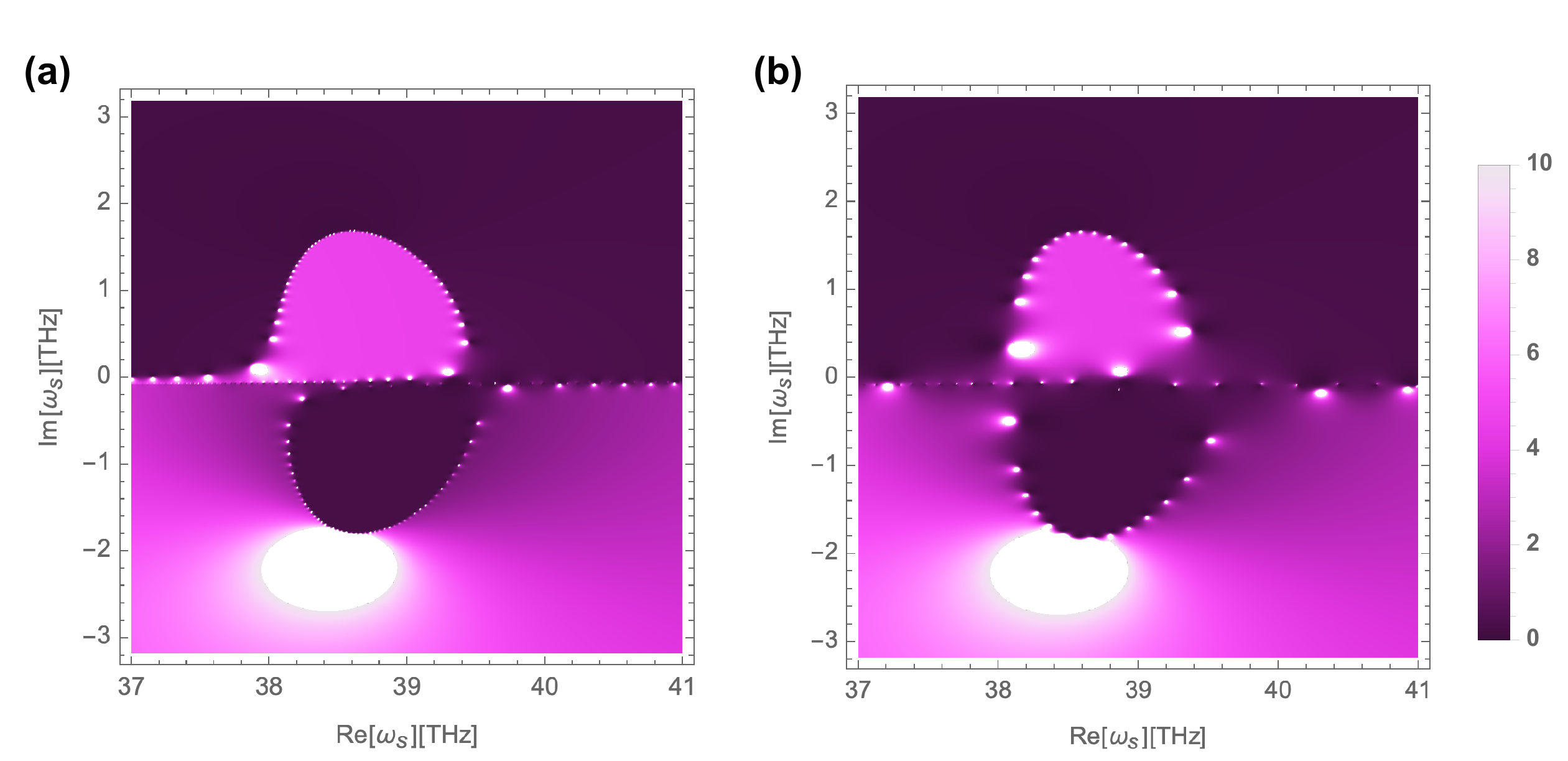}
\caption{(a) Reflectivity for a finite-thickness slab of $L=3\times10^{-3}$m. Small white circles indicate the presence of poles. The poles are located very close to the line of the discontinuity of $R_{\rm s}$ for the semi-infinite slab.
(b)	Same as the left figure for $L=9\times 10^{-4}$m. The density of poles is smaller than that for $L=3\times10^{-3}$m because it scales as $O(L)$.}
\label{fig Rfinite}
\end{figure*}

In Fig.~\ref{fig Rfinite} we present results of a more detailed analysis of the reflectivity in a slab of finite width of pumped {\rm SiC} 
at complex frequencies close to the frequency $\omega_{\rm s}$=39THz. 
This corresponds to one of the frequencies where we observed non-analytic behavior of $R_{\rm s}$ in the case of a semi-infinite slab. 
We observe that discrete poles are distributed on the circumference of a circle which extends into both upper and lower half-planes.   The circumference passes near the curves where $\real[\kappa]=0$ for either $\kappa_+$ or $\kappa_-$ and, in the limit of a semi-infinite sample, the circumference coincides exactly with these curves.
The poles on the upper-half of the circumference have positive imaginary frequency, {\em i.e.}, they correspond to the lasing modes. 
Discrete character of eigenmodes comes from using a slab of finite thickness, which gives an analogue of discrete modes in a cavity. 
As we take the slab thickness to infinity, the poles become denser and their residues are reduced. 
In the limit of a semi-infinite medium, discrete poles merge into a continuous line of poles with finite residue density per length. Then the reflectivity $R_{\rm s}$ develops a step-like discontinuity across the line of poles.

\begin{figure*}
	\includegraphics[width=2 \columnwidth]{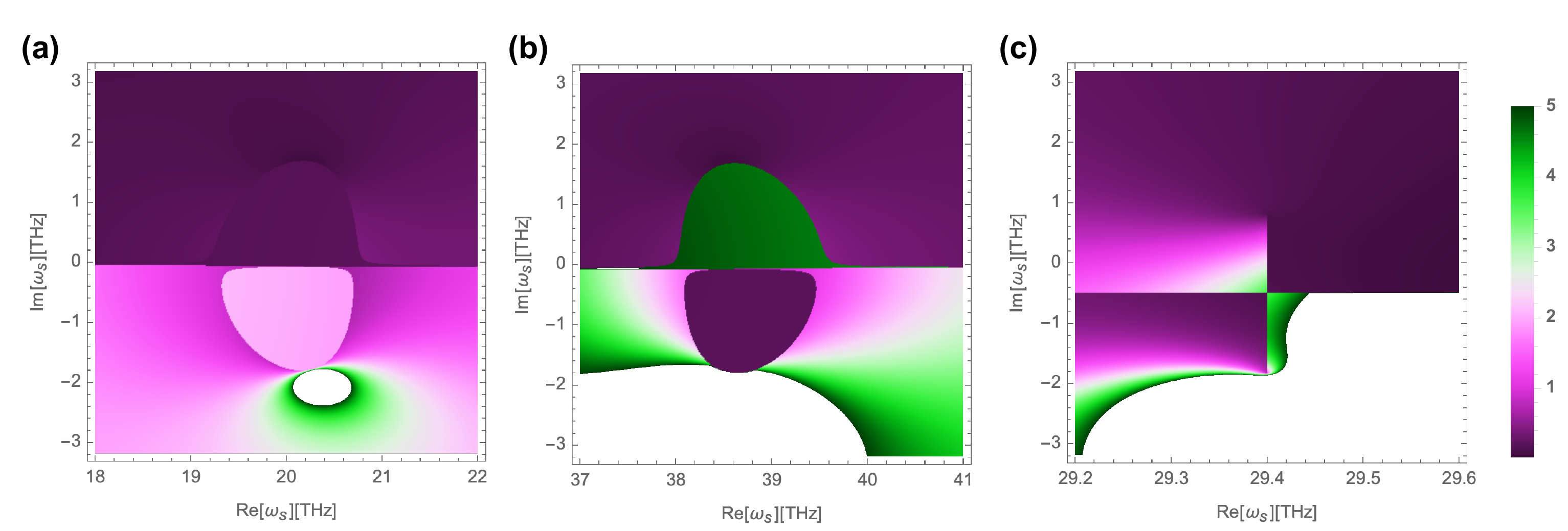}
\caption{(a) Reflectivity of the semi-infinite slab around $\omega_{\rm s}=$39THz. $R_{\rm s}$ is displayed in white when it is larger than 5. A circle-like line of the discontinuity of $R_{\rm s}$ appears at the center of the figure. It extends to positive $\img[\omega_{\rm s}]$. A straight line of discontinuity also exists close to the real axis, but it is located in the lower-half plane. The poles in the lower-half plane do not affect the electric field of the temporal problem \eqref{E_reflected_time_domain}.
	(b) Same as (a) for $\omega_{\rm s}$ around 20THz. A circle-like line of the discontinuity, which is symmetric to (a) appears.
	(c) Same as (a) for $\omega_{\rm s}$ around 29THz. We see a vertical line of discontinuity which extends to positive imaginary frequency.}
\label{fig Rinf}
\end{figure*}

Plots of $R_{\rm s}$ for an infinitely thick slab of {\rm SiC} as a function of complex frequency $\omega_{\rm s}$  are presented in Fig.~\ref{fig Rinf}. 
We also see that some poles are located near the real axis. 
However, they are not important in our analysis because closer examination shows that the line which satisfies $\real [\kappa]=0$ near the real axis is in the lower half-plane, indicating that the poles around this line have a finite lifetime. 

We observe a similar pattern of discontinuities in the imaginary part of $R_{\rm s}$ around frequencies  $\omega_{\rm s}=$20THz. This reflects the symmetry between $\omega_{\rm s}$ and $2\omega_{\rm p}-\omega_{\rm s}$ in $\kappa_\pm$.
On the other hand at $\omega_{\rm s}=\omega_{\rm p}$ we find poles not on a circle but on a straight line (see Fig.~\ref{fig Rinf}c). This can be understood by observing that symmetry $\kappa_+ = \kappa_-$ holds at this frequency.


\subsection{Enhanced reflection around 24 and 35 THz\label{Subsec SPhP}}
Examination of the reflection coefficient in the case without pumping (green line in figure \ref{reflectivity}) shows that it approaches one for frequencies
between 24 and 30THz. This frequency range corresponds to the reststrahlen band of SiC for which we do not find bulk polariton modes. In systems with pumping
we observe that in the same frequency range $\real [\kappa_{+}]$ is large (see figure \ref{fig kappas}d). This indicates that light incident at this frequency cannot penetrate deeply into the material. This is the remnant of the reststrahlen band in the static case. We also observe that $\real [\kappa_{+}]$ goes through the maximum 
at 24 THz. This is the frequency of the original phonon mode and corresponds to the strongest losses inside the medium. 
Due to the Floquet nature of the system $\real[\kappa_{+}]$ is also large for the ``Floquet mirror frequencies" 24 THz $< \omega_s < $ 35 THz. The largest value of $\real[\kappa_{+}]$
occurs close to 35 THz, which is a Floquet partner of the phonon frequency of 24 THz. Large values of $\real [\kappa_{+}]$ suggest that light cannot penetrate
the Floquet medium and we should find an increase in the reflection coefficient. This is indeed the case, as can be seen from Fig.~\ref{reflectivity}.

\begin{figure*}
     \includegraphics[width= 2 \columnwidth]{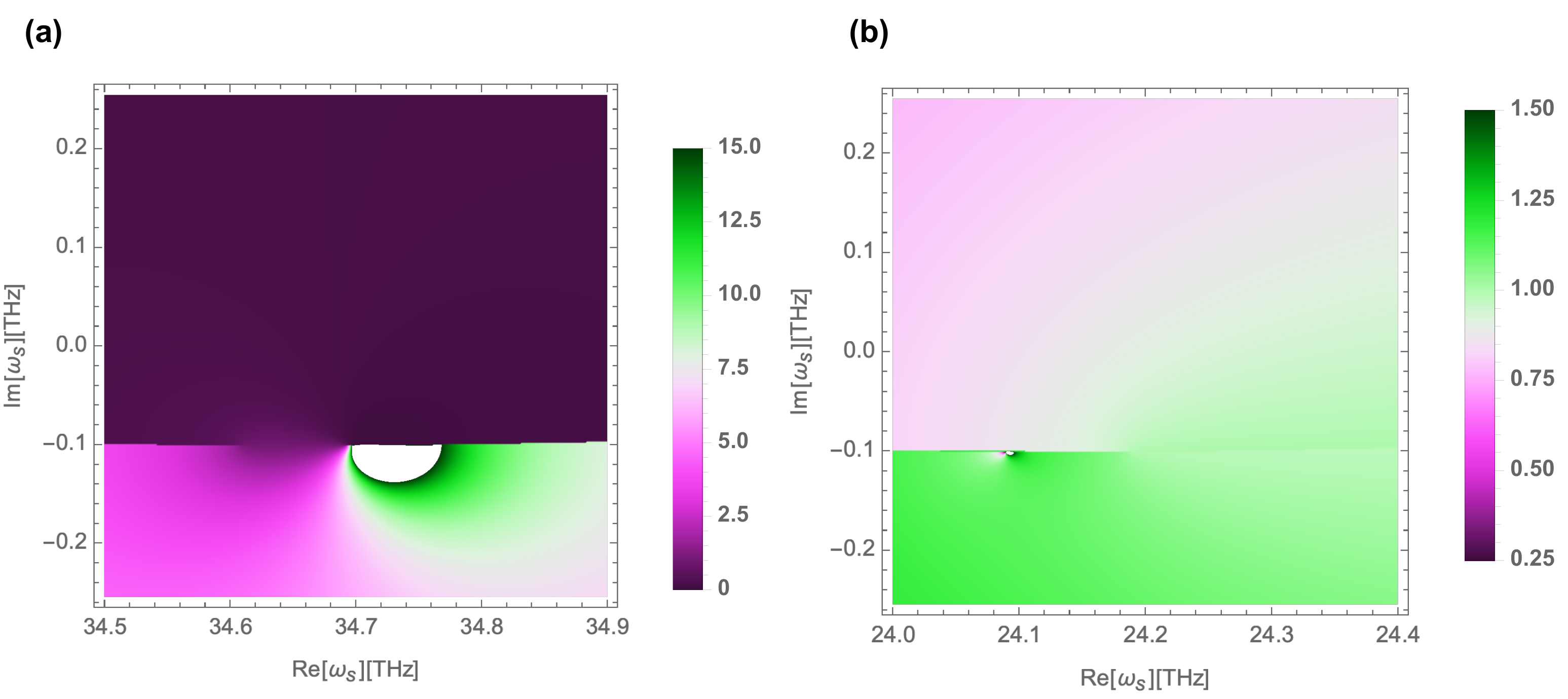}
    \caption{(a) Reflectivity $R_{\rm s}$ of the semi-infinite slab when the complementary frequency $\omega_{\rm i}=2\omega_{\rm p}-\omega_{\rm s}$ approaches the edge of the reststrahlen band. 
    $R_{\rm s}$ is displayed in white when it is larger than 15. There is a pole at $\omega_{\rm s}$=34.74THz. 
    (b) Same as (a) when the signal frequency $\omega_{\rm s}$ matches the edge of the reststrahlen band. There is a pole at $\omega_{\rm s}$=24.1THz.}
    \label{fig surfacemode}
\end{figure*}

It is interesting to point out that extending reflection coefficient to the complex $\omega_s$ we find isolated poles in the lower half-plane
with $\real[\omega_s]$ close to 24 and 35 THz and imaginary part slightly below $-\gamma/2$, where $\gamma$ is the phonon dumping coefficient, as shown in Fig. \ref{fig surfacemode}.
However we do not expect these poles to have any consequences for the reflection coefficient at real frequencies. These poles are separated
from the real axis by the branch cut in the reflection coefficient at $\img[\omega_s] = -\gamma/2$. Even in the static case, one finds isolated pole in the reflection coefficient in the lower half-plane with the real part of frequency matching the original phonon frequency and imaginary part slightly below $-\gamma/2$. In the static case one also has a branch-cut in the reflection coefficient at $\img[\omega_s] = -\gamma/2$ that separates these poles from the physical real axis.

\begin{table}[htb]
	\begin{tabular}{|l|c|} \hline
		$20$THz & lasing mode\\ \hline
		$24$THz & $\real[\kappa_{\pm}]$ is large, strong reflection  \\ \hline
		$29$THz & lasing mode\\ \hline
		$35$THz & $\real[\kappa_{\pm}]$ is large, strong reflection  \\ \hline
		$39$THz & lasing mode\\ \hline
	\end{tabular}
	\caption{Summary of the collective modes which correspond to the discontinuities and peaks of the reflectivity. }
	\label{table modes}
\end{table}

\section{Discussion of the parametric instability } \label{Sec eigenmodes}

\subsection{Lasing, causality, and afterglow}\label{Subsec lasing}

\begin{figure}[htp]
\begin{center}
	\includegraphics[width=0.4\textwidth]{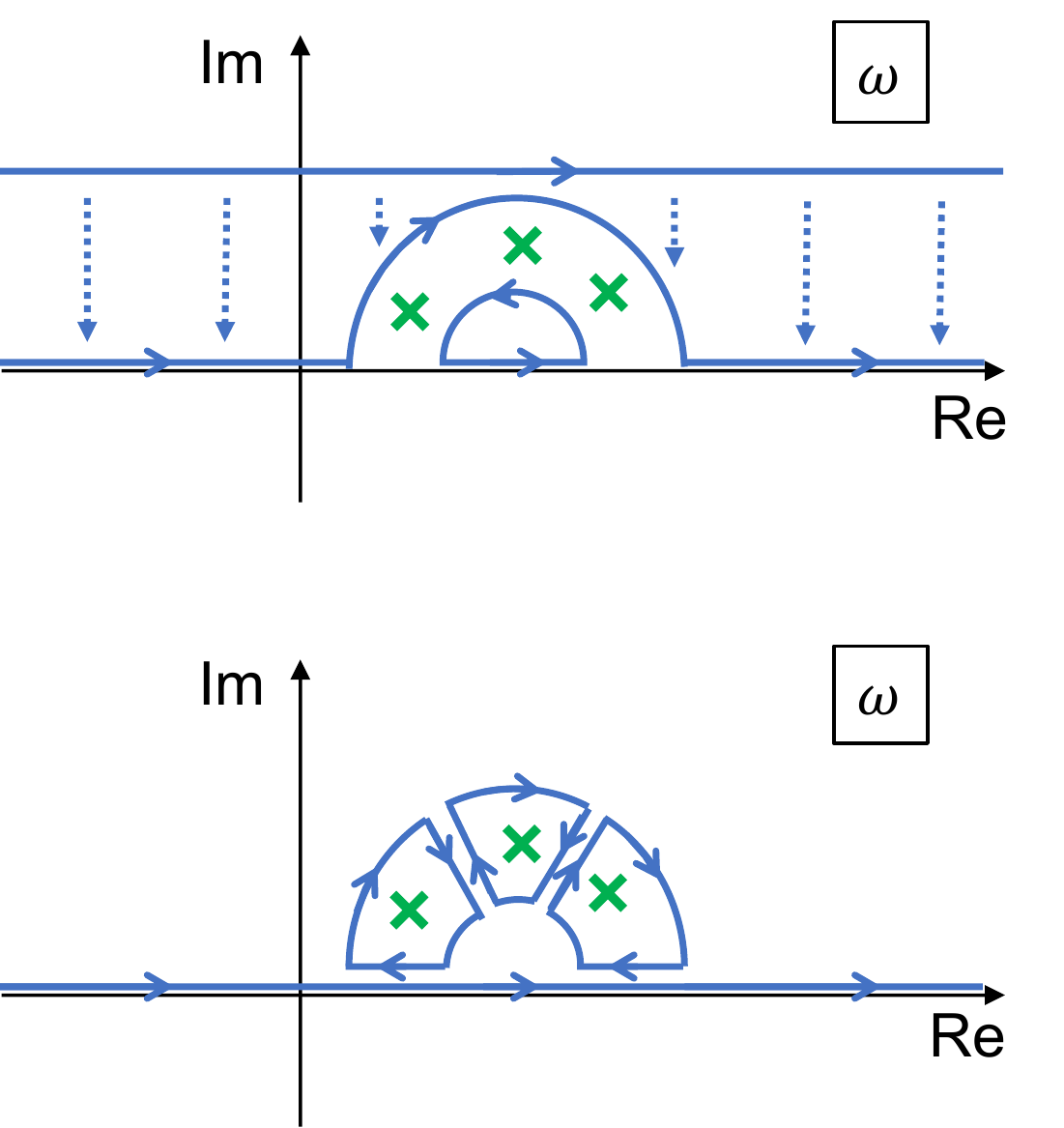}
	\caption{In the presence of unstable modes Fourier transforms should be defined on a contour that goes in the complex $\omega$ plane
	above all the poles of the reflection coefficient $r(\omega)$. This is equivalent to taking the integral over the real axis and adding poles corresponding to unstable
	modes. }
	\label{contour_unstable_poles}
\end{center}
\end{figure}

The existence of lasing modes modifies how one needs to analyze reflection of the probe pulse in the time domain. The standard approach is
to decompose the incoming pulse in Fourier series as
\begin{eqnarray}
e_0(z,\omega) = \frac{1}{2\pi} \int_{-\infty}^{+\infty} E_0(z,t) e^{i\omega t} dt
\end{eqnarray}
Then one uses scattering amplitude for individual Fourier components of $r_{\rm s}(\omega)$ and $r_{\rm i}(\omega)$ to construct the reflected wave as

\begin{eqnarray}
E_r(z,t)  
=  \int_{-\infty}^{+\infty} e_0(z,\omega) 
	\left[r_{\rm s}(\omega) e^{-i\omega t}+r_{\rm i}(\omega) e^{i(2\omega_{\rm p} - \omega) t} \right]
	d\omega
\label{E_reflected_time_domain}\non \\
\end{eqnarray}
As long as the poles of the reflection coefficient $r_{\{\rm s,i\}}(\omega)$ are in the lower half-plane one can show that this procedure satisfies causality:
reflected light does not appear before the incident pulse approaches the surface of the material. One can verify this by taking a pulse that
only hits the surface at $t=0$, {\em i.e.} for $t<0$ we have $E_0(z=0,t)=0$ and the function $e_0(z=0,\omega)$ is analytic in the upper half plane of complex $\omega$.
When computing the integral in (\ref{E_reflected_time_domain}) for $t<0$ one can close the contour of $\omega$ integration
in the upper half plane and prove that $E_r(z=0,t<0)=0$. When $r_{\{\rm s,i\}}(\omega)$ has poles in the upper half plane, this argument does not work and equation 
(\ref{E_reflected_time_domain}) formally implies nonzero amplitude of reflected light before the incident beam interacts with the material. To avoid this contradiction
one needs to define the range of frequency integration to be a line in the complex $\omega$ plane parallel to the real axis but going above
any poles of $r_{\{\rm s,i\}}(\omega)$ (see Fig.~\ref{contour_unstable_poles}). By deforming this contour as shown in Fig.~\ref{contour_unstable_poles}, we see that this is equivalent to taking a Fourier integral on the real axis and adding contributions from the poles of  $r_{\{\rm s,i\}}(\omega)$. Note that poles of $r_{\rm s}(\omega)$ and $r_{\rm i}(\omega)$ always coincide because the denominators of $r_{\rm s}(\omega)$ and $r_{\rm i}(\omega)$ are the same in \eqref{rs} and \eqref{r_i}. Then the temporal solution of the electric field is given by
\begin{align}
E_r(t) 
&= \int_{-\infty}^{+\infty} e_0(\omega) 
	\left\{r_{\rm s} e^{-i\omega t}+r_{\rm i} e^{i(2\omega_{\rm p}-\omega) t}\right\} d\omega
\nonumber\\
&-2\pi i \sum_j e_0(\omega_j) \left\{ 
		\underset{\omega_j}{\rm Res} (r_{\rm s}) e^{-i \omega_jt} 
		+\underset{\omega_j}{\rm Res} (r_{\rm i}) e^{i (2\omega_{\rm p}- \omega_j)t}
	\right\}
\label{E_reflected_time_domain2}
\end{align}
Taking $\omega_j = \omega_j'+ i \omega_j''$ we find that  for $\omega_j''>0$ the contribution coming from pole $j$ grows exponentially in time as $e^{\omega_j''t}$. 
At long times exponential growth of the reflected light should be limited by processes that we did not include in our analysis, such as non-linear interaction of
the signal and idler modes and their feedback depleting the pump field. 
However, we expect that our approach correctly predicts the characteristic time of the
``afterglow" that develops following the probe pulse after time
\begin{eqnarray}
\frac{1}{\tau} \sim {\rm max} \, \{ \, \omega_j'' \,\}
\end{eqnarray}
We note that even in the absence of the probe pulse, quantum and thermal noise may be sufficient to seed the parametric instability.
The ``afterglow"  should have Fourier components that match the frequency of unstable modes (see the $e^{i \omega_j' t}$ part of the second line of equation (\ref{E_reflected_time_domain2})) .
Hence the ``afterglow" signal should carry direct signatures of  unstable modes.
A similar residual response after a pulse is known as ringing \cite{Drever1983}. 
Ringing is the time response followed by an overshoot, resulting in decaying oscillations of increasing frequency \cite{Ying:07}.  
When we assume the phonon decays rapidly, the "afterglow" may exhibit a similar behavior to ringing. 
However, we also expect that the exponential growth of the reflected light will be observed in the presence of the strong pumped phonon-polariton.

Before concluding this section we would like to address the question of what makes frequencies of 20THz, 30 THz, and 39Thz special, so that  parametric instabilities take place
at these specific frequencies. Physical  mechanism responsible for parametric instabilities is shown in Fig. \ref{fig schematic}. Pumping creates a condensate of polaritons at the bottom
of the upper branch. Non-linearities present in the system allow a pair of polaritons from the condensate into a pair of excitations with opposite momenta  so that the total energy and momentum of the pair are conserved. 
Energies 20THz and 39Thz correspond to exactly such scattering where one of the
excitations stays in the upper branch and the other goes to the lower branch. And the 30THz instability corresponds to a scattering of a pair of polaritons  with a very small change of momenta and energy of individual excitations.
Similar mechanism of coherent scattering of bosonic excitations from a photoexcited coherent state has been observed earlier in experiments with exciton-polaritons \cite{Roberts1, Roberts2,Roberts3}.

\subsection{Lasing threshold and nature of unstable mode}\label{Subsec threshold}

\begin{figure}[h]
	\begin{center}
           \includegraphics[width=0.7\columnwidth]{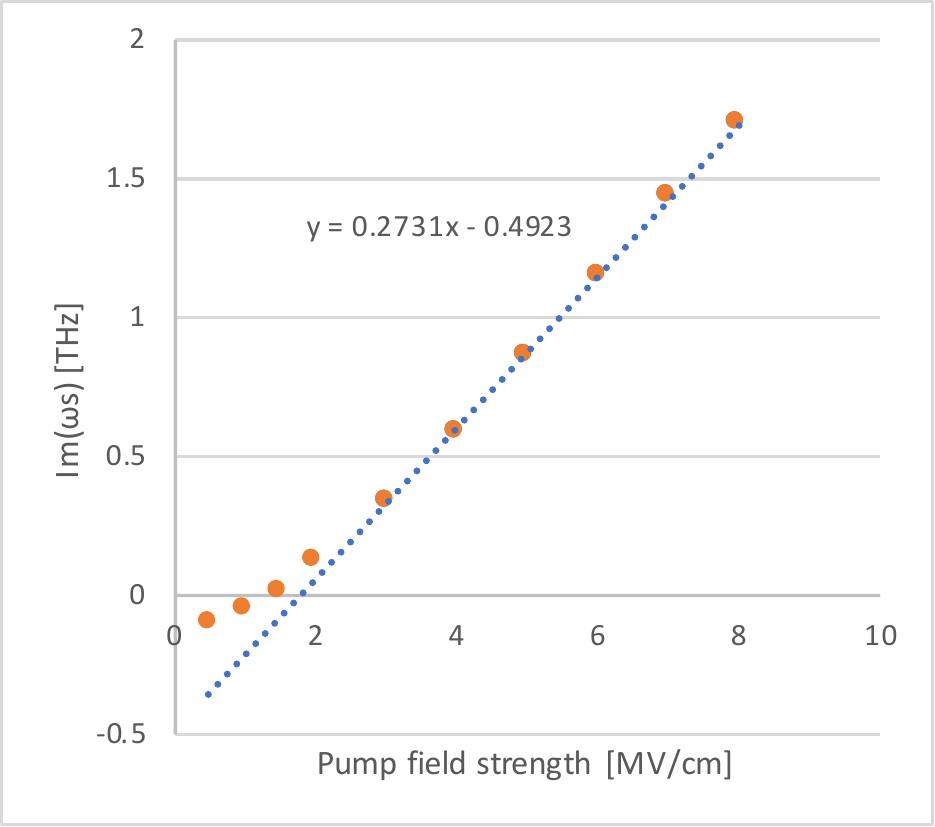}
    \caption{
    Growth rate of the most unstable lasing mode {\em vs} the strength of the pump field. 
    Beyond the threshold of $E_{\rm p}\simeq 1$MV/cm the instability growth rate increases linearly with $E_{\rm p}$} 
    \label{amp rate}
    \end{center}
\end{figure}
One interesting question related to lasing modes is whether they are present for infinitesimal amplitude of the pumped polariton amplitude or appear beyond a
certain threshold. In Fig.~\ref{amp rate}, we show the maximum value $\img[\omega_{\rm s}]$ of the unstable modes for different pump field strength $E_{\rm p}$. 
In equilibrium, {\em i.e.} when $E_{\rm p}=0$, all collective modes are in the lower-half plane. 
The first unstable mode appears for $E_{\rm p}=1$MV/cm. 
Since solutions for $\kappa_\pm$ are symmetric with respect to $\omega_{\rm s}$ and $\omega_{\rm i}$, unstable modes around frequencies $\omega_{\rm s}=$20THz and $\omega_{\rm s}=$39THz  appear simultaneously. 
Finite threshold for lasing in our system can be understood as a result of competition between parametric driving and losses in the medium.
Increasing $E_{\rm p}$ beyond the threshold value, we find that the maximum value of $\img[\omega_{\rm s}]$ increases linearly with amplitude of the pumped polariton.
Another family of unstable modes that we discussed before is at frequency $\omega_{\rm p}$. We find that the growth rate of this instability, {\em i.e.} the corresponding maximum value of $\img[\omega_{\rm s}]$, is always smaller than the growth rate of the modes at $\omega_{\rm s}=$20THz and $\omega_{\rm s}=$39THz. 
Note that the pump field strength used in Ref.~\cite{Cartella12148} is far beyond this threshold. 

\begin{figure}[h]
	\begin{center}
           \includegraphics[width=0.9\columnwidth]{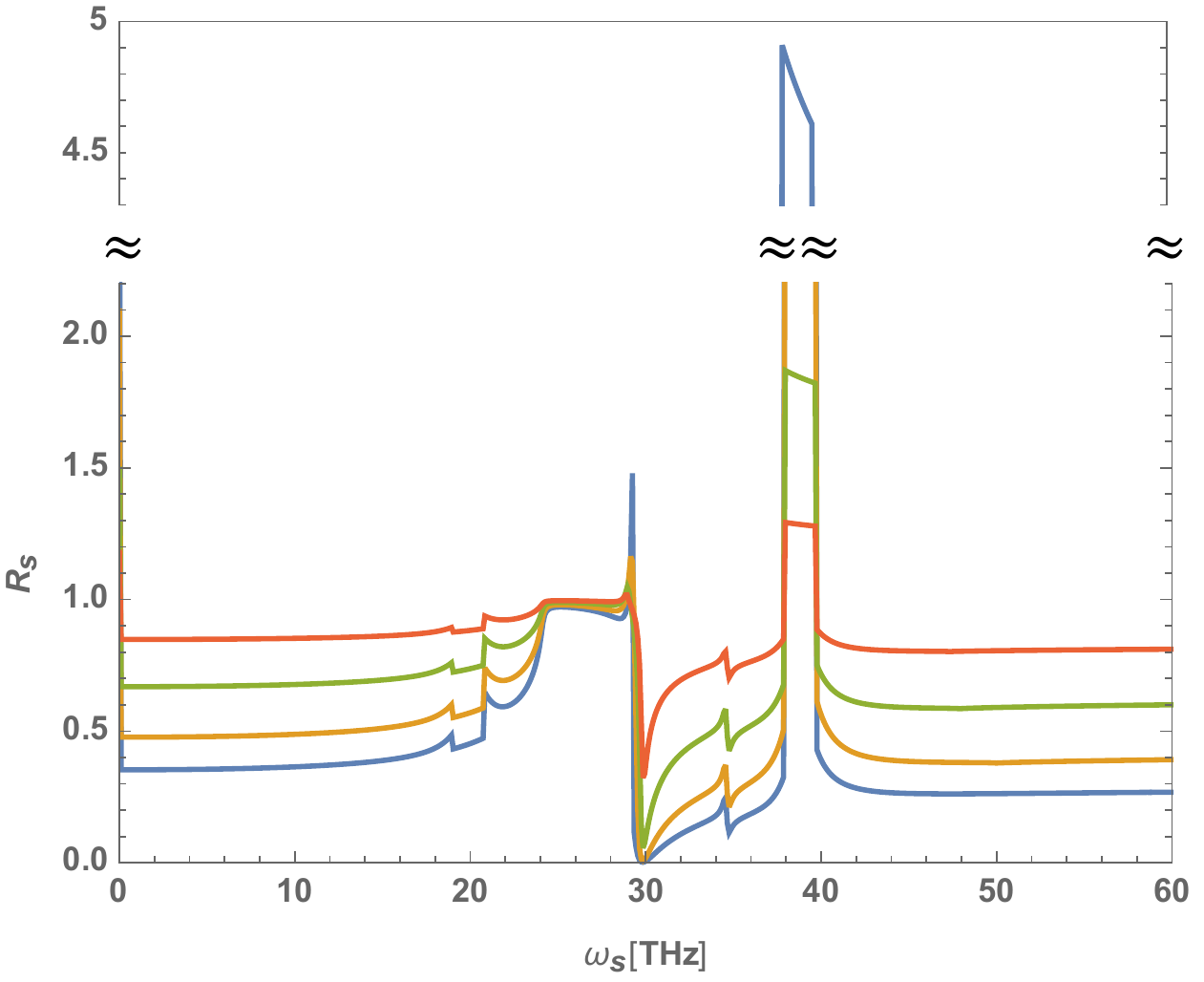}
    \caption{Angle dependence of the reflectivity. From the blue line to the purple line, lines corresponds to different angles of incidence: 0, $0.25\pi$, $0.375\pi$, and $0.45\pi$, respectively. Unstable modes at 20THz and 39THz are robust against the changes of the angle. By contrast, the unstable mode at 30THz turns into a surface mode and reflectivity becomes a continuous function of frequency for a non-normal incidence. 
}
    \label{fig nonnormal}
    \end{center}
\end{figure}

Another important test for the lasing mode is the robustness of discontinuities in the reflection coefficient with respect to changes in the direction of incident probe pulse.
Analysis of the angular dependence of discontinuities in the reflection coefficient also tells us about Floquet parametric instabilities for modes with finite momentum along the
surface of the material. 
In Fig.~\ref{fig nonnormal} we compare light reflection coefficients $R_s$ for different angles of incidence. At frequencies 20 and 39THz discontinuities remain albeit with a reduced magnitude. However, the discontinuity at 30THz disappears and is replaced by a sharp peak. 
To understand the persistence of the discontinuities at 20 and 39THz, 
we observe that Floquet eigenmode \eqref{kappa pm} does not depend on the direction of light inside the medium. Therefore, for any incidence angle, the phase matching condition in \eqref{Kappas_Kappai_condition} are satisfied at these frequencies and the eigenmode is infinitely extended inside the slab. 
From the perspective of two polariton scattering shown in Fig.~\ref{fig schematic}, it is possible to find a resonant scattering process that involves two polariton bands when the scattered photons have a fixed momentum along one of the axes. 
On the other hand, the instability arising from the small-momentum resonant polariton scattering within the upper polariton band does not exist when the scattered phonons have finite component along one of the axes. Hence incident light cannot penetrate the surface of the slab and the discontinuity at 30THz for normal incidence becomes a peak for finite incidence angle. 
In real experiments incident beams strike the surface with a finite range of incident angles. In light of the earlier discussion we expect that discontinuities in the reflection coefficients $r_{\rm s}$ and $r_{\rm i}$ at frequencies 20 and 39THz should be more readily observed in experiments. 
Our result in Fig.~\ref{fig nonnormal} is for S-polarized wave incidence. Regarding the P-polarized case, we note that the effective potential \eqref{Hamiltonian nonlinear} has cubic terms, which change the result quantitatively. However, the scattering process at 20 and 39THz should be robust as far as fourth order terms are present in the effective potential. 

\begin{figure}[h]
	\begin{center}
           \includegraphics[width=0.9\columnwidth]{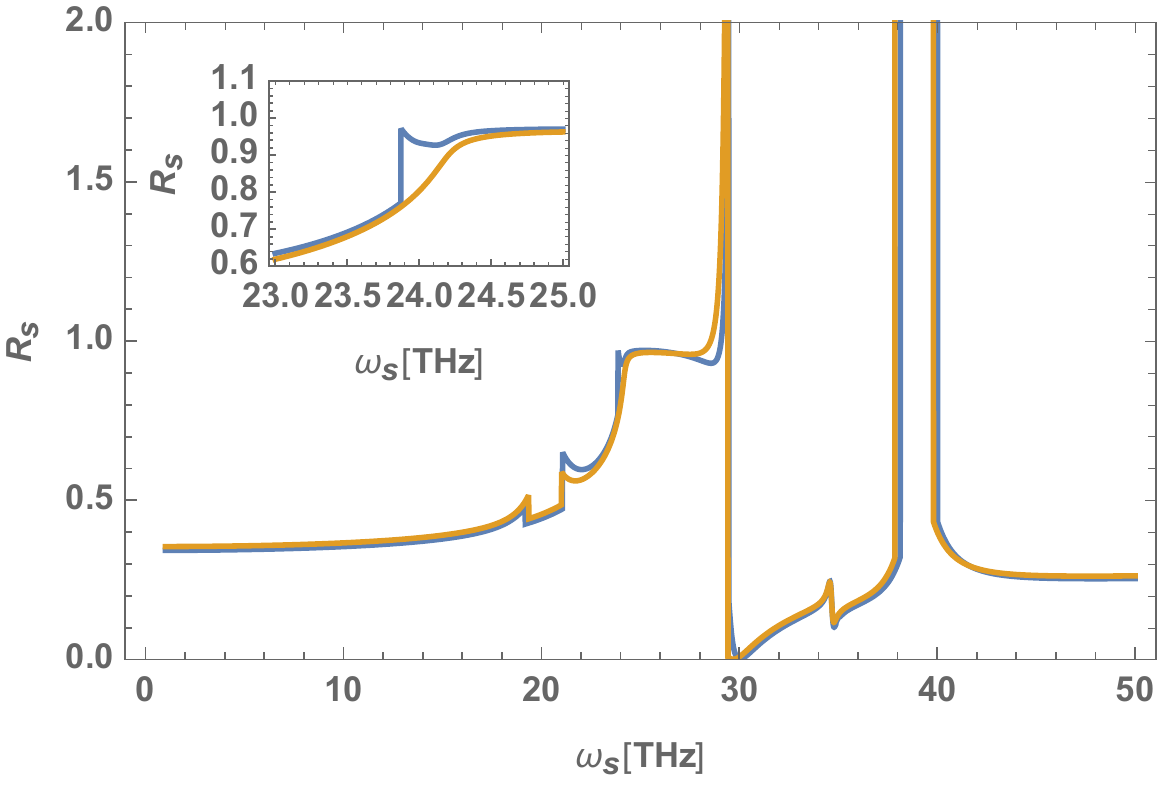}
    \caption{Reflectivity including the sum frequency component. The blue/orange line is with/without the sum frequency component. Although the reflectivity behaves qualitatively similar with the sum frequency component, the peaks at 24 and 35THz are slightly enhanced. 
 		Furthermore, another discontinuity appears at 23.8 THz.
 		 (Inset) Close-up of the discontinuity which appears when we include the third mode in our analysis:  $\omega_{\rm {sum}}=2\omega_p+\omega_s$ . 
}
    \label{three wave rs}
    \end{center}
\end{figure}

\begin{figure}[h]
	\begin{center}
           \includegraphics[width=0.9\columnwidth]{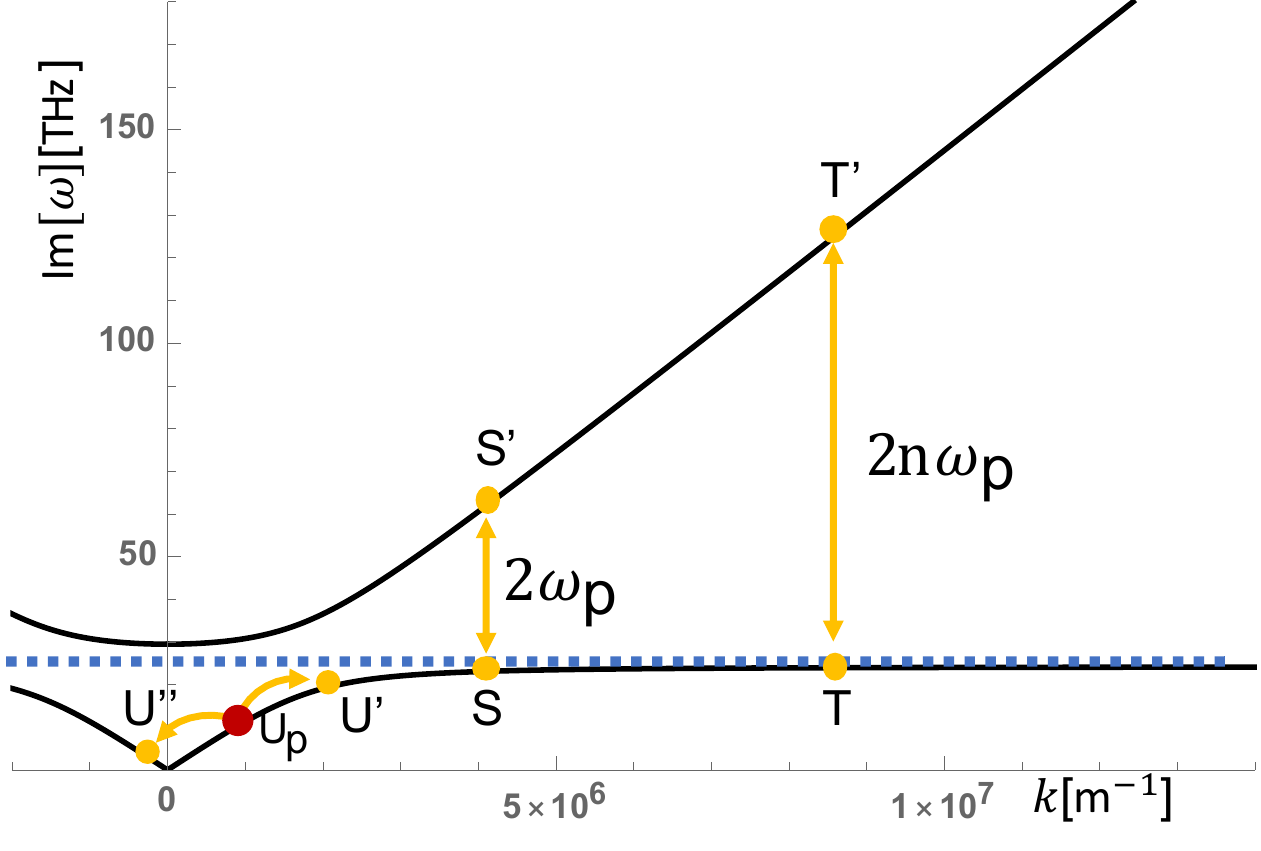}
    \caption{Resonant process corresponding to $\img [\kappa_s] = \img [\kappa_{\rm {sum}}]$. This can be understood as a four polariton process of two polaritons from the bottom of the upper polariton band and one polariton from the point S on the lower polariton band combining to make one polariton in the upper band point S'. Higher order processes of polariton scattering from T to T' offer absorbing energy $2n \omega_{\rm p}$ are also possible. 
    	In the case of the polaritons pumped at finite momentum, the process in which two polaritons scatter from ${\rm U}_{\rm p}$ to U' and U'' is also possible. 
}
    \label{fig other scattering}
    \end{center}
\end{figure}

Before concluding this section we will discuss effects of including more modes, arising from other harmonics in the Floquet analysis. In particular we observe that to linear order in $E_{\rm p}^2$ we have mixing between $\omega_{\rm s}$ and frequencies $\omega_{\rm i}=2\omega_{\rm p}-\omega_{\rm s}$ and $\omega_{\rm sum}\equiv 2\omega_{\rm p}+\omega_{\rm s}$. 
So far our discussion focused on the role of $\omega_{\rm i}$. We can extend our analysis to include $\omega_{\rm sum}$. Results of the numerical solution of the model with three coupled frequencies is shown in Fig.~\ref{three wave rs}. We find few differences with the solution of the two frequencies case. Generally the wave vector corresponding to $\omega_{\rm sum}$ does not match the wavevector corresponding to $\omega_{\rm s}$. Thus hybridization between $\omega_{\rm s}$ and $\omega_{\rm sum}$ is strongly suppressed. One exception is the points shown in Fig.\ref{fig other scattering}. 
Closer examination of Fig.~\ref{three wave rs} shows that indeed for the frequency $\omega_{\rm s}=23.8$THz analysis that includes three mode mixing exhibits a small discontinuity of $R_{\rm s}$. The origin of this discontinuity is a resonant excitation from the LP branch point S to the UP point S' (see fig. \ref{fig other scattering}) by the Floquet modulation of the medium (coherent oscillations excited by the pump pulse).
Generalization of this argument suggests that we can also find higher order resonances when states in the upper and lower bands at the same momentum differ in energy by $2n\omega_{\rm p}$. All of them are expected to exhibit lasing, although they involve higher non-linearities and therefore require a larger threshold pump intensity for lasing to occur. 

\section{Conclusions}\label{Sec Conclusions}
In this paper we developed an analytic theory for analyzing  pump and probe experiments in a polariton system with 
resonant pumping of the upper polariton mode. We computed the response of the system to the probe pulse 
by solving the coupled set of phonon and Maxwell equations linearized around the coherently oscillating upper polariton mode. Our method utilizes the Floquet approach in which pump-induced coherent polariton oscillations provide time-periodic modulation of the
medium properties. The key feature of such Floquet-Fresnel light reflection analysis is frequency mixing analogous to that of parametric amplifiers. The probe light incident at frequency $\omega_{\rm s}$ results in the appearance of the reflected and transmitted waves at frequency $\omega_{\rm i} = 2 \omega_{\rm p} - \omega_{\rm s}$, where $\omega_{\rm p}$ is the frequency of the polariton mode excited by the pump. In the context of non-linear optics such complementary frequency modes are usually referred to as idler modes. Earlier experiments in SiC demonstrated light amplification of the probe beam upon reflection, which in our analysis can be understood as originating
from photon pairs produced by the oscillating polariton mode. We find that when the
pump induced polariton amplitude is strong enough, the system should not only exhibit amplification but undergo a true parametric instability. Such instability can be understood as arising from the two polariton 
scattering process that satisfies energy and momentum conservation (see Fig.~\ref{fig schematic}). This instability mechanism is analogous to stimulated resonant scattering of exciton-polaritons in the case of pumping at the "magic angle" demonstrated by Baumberg et al. \cite{Roberts3}.
While earlier pump and probe experiments with polaritons in SiC were performed using broad band probe pulses, we showed that using frequency resolved probes should provide new insights into properties of the photoexcited system. In particular reflection
coefficients for both  the signal and idler components should exhibit discontinuities in the frequency dependence arising from the parametric instability. 
We note however that true discontinuities are present only in the idealized situation of a uniformly excited infinitely large crystal and infinitesimal intensity of the probe beam. Under realistic experimental conditions of finite penetration depth of the pump pulse and small but finite intensity of the probe pulse, discontinuities in the reflection coefficient should become sharp crossovers.
We discuss the possibility of the ``afterglow" that can appear after the probe pulse is gone, which can be understood as the parametric instability triggered by the probe pulse. Another interesting feature in the frequency dependence of the reflection coefficients that we discuss are sharp peaks arising from the surface modes that appear near the edge of the reststrahlen band for both the signal and idler frequency components. 

In addition, another scattering process may be realized when we pump polaritons at finite momentum in the lower polariton branch. 
In Fig.~\ref{fig other scattering} we show the situation when the pumped polaritons have a macroscopic occupation at ${\rm U}_{\rm p}$ and scatter to U' and U''. 
This is a direct analogue of the exciton resonant scattering observed by  Baumberg et al. \cite{Roberts3}.
In this process the scattering occurs in the lower branch, in contrast to the situation shown in Fig.\ref{fig schematic}, where
the polaritons are scattered into both the upper and lower branches.
Our Floquet formalism can be extended to study the optical properties of the system in the presence of polaritons pumped at finite momentum. 

We envision several potential technological applications of the system discussed in this paper. Lasing instability allows to obtain coherent sources at 20 and 40 THz starting from radiation at 30 THz. Finding strong sources at lower frequencies is particularly valuable in mid-IR domain. Moreover we expect that the parametric frequency can be tunable. In this paper we analyzed the simplest situation of pumping at zero wavevector. One can extend Floquet-Fresnel analysis to the situation when the pump excites upper polaritons at finite momentum. This will modify resonant scattering process shown in Fig.~\ref{fig schematic}  and should result in the modification of lasing frequencies. We also note that resonant polariton scattering discussed in our paper results in the production of entangled THz photons, which may find applications in quantum information. Another promising direction is to consider pumping regime when the system is about to develop parametric instability, so the poles move from the lower to the upper half-plane. At the transition we should find polariton modes at real frequencies which indicates that they can propagate in the material for a long time without getting damped and becoming unstable. This is the regime of exact balance between dissipation and Floquet pumping. Such modes can be used for efficient frequency filtering.
By changing the momentum of the pump photons, we expect to be able to modify the frequency
that can be filtered. 

Discussion presented in our paper considered a specific type of optical non-linearity. We expect that mechanism of Floquet parametric instability that we discussed here may be present in a broad class of systems. When a phonon (or two phonons) can scatter into a pair of collective modes such that energy and momentum are satisfied, we expect to find Floquet parametric instabilities.

\section*{Acknowledgement}

The authors thank A.~Cartella, A.~Cavalleri, J.~Faist, B.~Halperin, A.~Imamoglu, A.~Rosch, for useful discussions and valuable comments. We thank A. Leitenstorfer for a careful reading of the manuscript and insightful comments. We acknowledge support  from  the Harvard-MIT CUA, Harvard-MPQ Center, 
AFOSR-MURI: Photonic Quantum Matter (award FA95501610323), DARPA DRINQS program (award D18AC00014), the Israel Science Foundation (grant 1803/18), and the National Science Foundation through a grant to ITAMP at the Harvard-Smithsonian Center for Astrophysics. 
SS acknowledges support from JSPS Overseas Research Fellowships.
DP thanks the Aspen Center for Physics, where part of this work was done, with support of NSF grant number PHY-1066293.

\appendix
\section{Nonlinear polariton analysis}

The nonlinear dynamics of the coupled phonon amplitude and electric field in SiC is described by equations (\ref{EOM Q}) and (\ref{Maxwell combined}).   We begin by linearizing equation (\ref{EOM Q}) with respect to the signal and idler frequency components 
\begin{align}
	g(\omega_{\rm s})Q_{\rm s}
	=&
	\eta E_{\rm s} + 2\beta Q_{\rm p}^2 (-2\nu E_{\rm i}^*+\nu^2 Q_{\rm i}^*) \non \\
	&+2\beta |Q_{\rm p}|^2 (-4\nu E_{\rm s} + 2 \nu^2 Q_{\rm s}) \non\\
	&+3\alpha Q_{\rm p}^2(E_{\rm i}^* -2 \nu Q_{\rm i}^*) +3\alpha |Q_{\rm p}|^2(2E_{\rm s}-4\nu Q_{\rm s}),
	\label{gQs1}
\end{align}
where we used the relation \eqref{Ep Qp nu} to express $E_p=\nu Q_p$. 
Separating $Q_{\rm s}$ and $E_{\rm s}$ terms of the last equation, we obtain
\begin{align}
&\tilde{g}(\omega_{\rm s})Q_{\rm s} \non \\
&=
\left[\eta +(-8\beta \nu +6\alpha )|Q_{\rm p}|^2 \right]E_{\rm s} 
+(-4\beta \nu +3\alpha)Q_{\rm p}^2E_{\rm i}^* \non \\
&+(2\beta \nu^2-6\alpha \nu)Q_{\rm p}^2 Q_{\rm i}^*.
\label{gQs2}
\end{align}
where 
\begin{align}
\tilde{g}(\omega) \equiv g(\omega)+(-4\beta\nu^2 + 12 \alpha \nu)|Q_{\rm p}|^2
\end{align}
We write analogous equations for the idler frequency component and solve for the two phonon
amplitudes, $Q_s$ and $Q_i^*$ in terms of the electric fields $E_s$ and $E_i^*$.   Assuming that the polariton amplitude  $Q_{\rm p}$ is not too large, we keep terms up to quadratic order in  $Q_{\rm p}$.
We find 
\begin{align}
	Q_{\rm s}=& 
	{1\over \tilde{g}(\omega_{\rm s})}\left[
		\eta+(-8\beta \nu +6\alpha) |Q_{\rm p}|^2
		\right]E_{\rm s} \non \\
	+&{1\over \tilde{g}(\omega_{\rm s})} \left[ 
		(-4\beta \nu + 3\alpha ) 
		+{\eta(2\beta \nu -6\alpha \nu) \over \tilde{g}(-\omega_{\rm i})}
	\right] Q_{\rm p}^2 E_{\rm i}^*.
	\label{Qs solution Appendix}
\end{align}
and
\begin{align}
	Q_{\rm i}^* =& 
	{1\over \tilde{g}(-\omega_{\rm i})}\left[
		\eta+(-8\beta \nu +6\alpha) |Q_{\rm p}|^2
		\right]E_{\rm i}^* \non \\
	+&{1\over g(-\omega_{\rm i})} \left[ 
		(-4\beta \nu + 3\alpha ) 
		+{\eta(2\beta \nu -6\alpha \nu) \over g(\omega_{\rm s})}
	\right] Q_{\rm p}^{*2} E_{\rm s}.
	\label{Qi solution Appendix}	
\end{align}

We now consider the electric field equation \eqref{Maxwell combined} and separate the $\omega_s$ frequency component of this equation. 
\begin{widetext}
\begin{align}
	k^2 E_{\rm s} -{\omega_{\rm s}^2 \over c^2}\Bigg[
		\left(\epsilon_\infty + {4\beta \over \epsilon_0}|Q_{\rm p}|^2\right)E_{\rm s} 
		+\left\{{\eta \over \epsilon_0}+\left(-8\beta \nu +6\alpha \over \epsilon_0 \right) |Q_{\rm p}|^2 \right\}Q_{\rm s} 
		+\left({3\alpha - 4\beta \nu \over \epsilon_0} \right) Q_{\rm p}^2 Q_{\rm i}^*
		+{2 \beta \over \epsilon_0} Q_{\rm p}^2 E_{\rm i}^*
	\Bigg]=0,
\end{align}
\end{widetext}
where $k\equiv |{\bf k}|$. We use \eqref{Qs solution Appendix} and \eqref{Qi solution Appendix} to eliminate the phonon amplitudes $Q_s$ and $Q_i^*$ and obtain
\begin{align}
	k^2 E_{\rm s} 
	-{\omega_{\rm s}^2 \over c^2}\left(
		\tilde{\epsilon}(\omega_{\rm s}) E_{\rm s}
		+G(\omega_{\rm s}) Q_{\rm p}^2 E_{\rm i}^*
	\right)=0,
	\label{Maxwell Es Appendix}
\end{align}
where 
\begin{align}
	\tilde{\epsilon}(\omega)
		\equiv& \epsilon_\infty + {\eta^2 \over \epsilon_0 \tilde{g}(\omega)}  + F(\omega_{\rm s}) |Q_{\rm p}|^2  
		\label{tilde epsilon Appendix} \\
		F(\omega) \equiv& {1\over \epsilon_0}\left[
			4\beta + {4\eta\over \tilde{g}(\omega)}(3\alpha-4\beta \nu)
		\right]\\
G(\omega) 
\equiv &
{1 \over \epsilon_0}\Bigg[
2\beta+\left({\eta \over \tilde{g}(\omega)}+{\eta \over \tilde{g}(\omega-2\omega_{\rm p})}\right)(3\alpha -4 \beta \nu ) \non \\
&+{2\eta^2 \over \tilde{g}(\omega) \tilde{g}(\omega-2\omega_{\rm p})} (-3\alpha \nu + \beta \nu^2)
\Bigg]
\end{align}
Repeating the same calculation for the $\omega_{\rm i}$ frequency component, we obtain
\begin{align}
	k^2 E_{\rm i}^* 
	-{\omega_{\rm i}^2 \over c^2}\left(
		\tilde{\epsilon}(-\omega_{\rm i}) E_{\rm i}^*
		+G(\omega_{\rm s}) Q_{\rm p}^{*2} E_{\rm s}
	\right)=0
	\label{Maxwell Ei Appendix}
\end{align}
\eqref{Maxwell Es Appendix} and \eqref{Maxwell Ei Appendix} are linear equations which describe the propagation of light inside the Floquet medium. 


\section{ Analysis of light scattering from a slab of finite thickness} \label{Sec Finite thickness}

In this section we present analysis of light reflection from a slab
of finite thickness in the case of normal angle of incidence. We set the boundaries 
of the material to be at $z=0$ and $z=-L$. 
In the air for $z>0$, we use \eqref{reflection incident} and \eqref{EB ref}. 
In the insulator for $-L<z<0$
\begin{eqnarray}
\left( \begin{array}{c} {\bf E}(z,t) \\ {\bf B}(z,t) \end{array} \right)
&=& \left( \begin{array}{c}  {E}_{\rm slab}(z,t) \hat{e}_x \\ {B}_{\rm slab}(z,t) \hat{e}_y \end{array} \right)
\nonumber
\\
\left( \begin{array}{c}  {E}_{\rm slab}(z,t) \\ {B}_{\rm slab}(z,t)  \end{array} \right)
&\equiv& t_1 \, \left( \begin{array}{c}  {E}_{+}(t)  \\ {B}_{+}(t) \end{array} \right)
e^{ \kappa_+ z}
\nonumber\\
&+&
t_2 \, \left( \begin{array}{c}  {E}_{+}(t)  \\ -{B}_{+}(t)  \end{array} \right)
e^{ -\kappa_+ z} \non \\
&+&
t_3 \, \left( \begin{array}{c}  {E}_{-}(t)  \\ {B}_{-}(t)  \end{array} \right)
e^{ \kappa_- z}\non \\
&+&
t_4 \, \left( \begin{array}{c}  {E}_{-}(t)  \\ -{B}_{-}(t)  \end{array} \right)
e^{ -\kappa_- z}.
\label{reflection slab finite}
\end{eqnarray}
In the air for $z<-L$, 
\begin{eqnarray}
\left( \begin{array}{c} {\bf E}(z,t) \\ {\bf B}(z,t) \end{array} \right)
&=& \left( \begin{array}{c}  {E}_{\rm trans}(z,t) \hat{e}_x \\ {B}_{\rm trans}(z,t) \hat{e}_y \end{array} \right)
\nonumber
\\
\left( \begin{array}{c}  {E}_{\rm trans}(z,t) \\ {B}_{\rm trans}(z,t)  \end{array} \right)
&\equiv& t_{\rm s} E_0 \, \left( \begin{array}{c}  1 \\ \sqrt{\epsilon_0 \mu_0} \end{array} \right)\, e^{ -i \omega_{\rm s} t} e^{ \kappa_{\rm s} z}
\nonumber\\
&+ & t_{\rm i} E_0\, \left( \begin{array}{c}  1 \\ \sqrt{\epsilon_0 \mu_0} \end{array} \right)\, e^{ i \omega_{\rm i} t} e^{\kappa_{\rm i} z} \label{EB ref finite}
\end{eqnarray}
In Sec.~\ref{Sec reflection}, we discussed the case of a semi-infinite slab and included only evanescent waves $e^{\kappa_\pm z}$ that vanish as $z \rightarrow - \infty $. In the case of a slab with finite thickness we need to include solutions corresponding to evanescent waves propagating from the lower interface, which correspond to the $e^{-\kappa_{\pm} z}$ terms in equation (\ref{EB ref finite}). The reflection and transmission coefficients can now be obtained by analyzing boundary conditions for the Maxwell equations at $z=0$ and $z=-L$. 
We have for $z=0$ 
\begin{align}
&1+r_{\rm s} =t_1+t_2+t_3+t_4 \label{finite 1}\\
&\kappa_{\rm s}(1-r_{\rm s}) =\kappa_+(t_1-t_2)+\kappa_-(t_3-t_4)\\
&r_{\rm i} =\alpha_+(t_1+t_2)+\alpha_-(t_3+t_4) \\
&-\kappa_{\rm i} r_{\rm i} =\kappa_+\alpha_+(t_1-t_2)+\kappa_-\alpha_-(t_3-t_4),
\end{align}
and for $z=-L$ 
\begin{align}
&t_1' +t_2'+t_3'+t_4'  =t_{\rm s} \\
& \kappa_+(t_1'-t_2')+\kappa_-(t_3'-t_4')=-\kappa_{\rm s} t_{\rm s}\\
&\alpha_+(t_1'+t_2')+\alpha_-(t_3'+t_4') =t_{\rm i} \\
&\kappa_+\alpha_+(t_1'-t_2')+\kappa_-\alpha_-(t_3'-t_4')=-\kappa_{\rm i} r_{\rm i},\label{finite 8}
\end{align}
where
\begin{align}
&t_1'\equiv t_1 e^{-\kappa_+ L} \label{t1'}\\
&t_2'\equiv t_2e^{\kappa_+ L}\label{t2'}\\
&t_3'\equiv t_3e^{-\kappa_- L}\label{t3'}\\
&t_4'\equiv t_4e^{\kappa_- L}\label{t4'}.
\end{align}
We assume $\real[\kappa_+]<\real[\kappa_-]$ without loss of generality. 
Then, the exponents in \eqref{t1'}-\eqref{t4'} satisfy 
\begin{align}
e^{-\kappa_- L} \ll e^{-\kappa_+ L} \leq e^{\kappa_+ L} \ll e^{\kappa_- L}. 
\end{align}
Here the equation in the middle holds when $\real[\kappa_+]=0$. 
From these relations one finds that $t_3'$ and $t_4$ are exponentially small, which allows
us to simplify the analysis by taking $t_3'=0$ and $t_4=0$. We can now write equations \eqref{finite 1}-\eqref{finite 8} as 
\begin{align}
&1+r_{\rm s} =t_1+t_2+t_3\\
&\kappa_{\rm s}(1-r_{\rm s}) =\kappa_+(t_1-t_2)+\kappa_-t_3\\
&r_{\rm i} =\alpha_+(t_1+t_2)+\alpha_- t_3 \\
&-\kappa_{\rm i} r_{\rm i} =\kappa_+\alpha_+(t_1-t_2)+\kappa_-\alpha_- t_3,
\end{align}
at $z=0$, and 
\begin{align}
&t_1' +t_2'+t_4'  =t_{\rm s} \\
& \kappa_+(t_1'-t_2')-\kappa_- t_4'=-\kappa_{\rm s} t_{\rm s}\\
&\alpha_+(t_1'+t_2')+\alpha_- t_4' =t_{\rm i} \\
&\kappa_+\alpha_+(t_1'-t_2')-\kappa_-\alpha_- t_4'=-\kappa_{\rm i} r_{\rm i},
\end{align}
at $z=-L$.
Solving these equations, we obtain $r_{\rm s}$. 
We do not present explicit expression for $r_{\rm s}$ because it is somewhat cumbersome and is not needed for our discussion. Instead we discuss the condition for $r_{\rm s}$ to have a pole, which corresponds to a polariton eigenmode of the system. 
The condition for the pole of $r_{\rm s}$ to be zero is given by the equation 
\begin{align}
e^{2\kappa_+ L} 
= A \label{exp 2kp}
\end{align}
Here A is the coefficient determined by the boundary conditions that can be written as
\begin{widetext}
\begin{align}
A\equiv
{
\{\alpha_+ (\kappa_i - \kappa_+) (\kappa_- - \kappa_s) - 
\alpha_- (\kappa_i - \kappa_-) (\kappa_+ - \kappa_s)\} 
\{\alpha_+ 
(\kappa_i + \kappa_+) (\kappa_- + \kappa_s) - \alpha_- 
(\kappa_i + \kappa_-) (\kappa_+ + \kappa_s)\}
\over
\{\alpha_- (\kappa_+ - \kappa_s)(\kappa_i + \kappa_-)  + \alpha_+ (\kappa_i 
- \kappa_+) (\kappa_- + \kappa_s)\} 
\{\alpha_+ (\kappa_i + 
\kappa_+) (\kappa_- - \kappa_s) + \alpha_- (\kappa_i - 
\kappa_-) (\kappa_+ + \kappa_s)\}
} .
\label{exp 2kp}
\end{align} 
\end{widetext}
Note that $A$ is a function of $\kappa$ but is independent of $L$. 
In all cases relevant to our discussion $A$ is finite, so we can take the logarithm of equation \eqref{exp 2kp} and find the eigenmode conditions 
\begin{align}
	\real[\kappa_+] &= {\log|A|\over L}
	\label{condition poles finite re}\\
	\img[\kappa_+] &= {n \over L}\pi + {\arg(A)\over L}, 
	\label{condition poles finite im}
\end{align}
where $n$ is an arbitrary integer. 

Physically eigenmodes correspond to waves that undergo infinite number of reflections between the upper and lower boundaries of the slab. Equation (\ref{condition poles finite re}) describes the balance between loss/gain during propagation in the slab coming from the dissipation/Floquet pumping ($\real[\kappa_+]L$) and loss during the reflection at the interface with air ($A$). Equation (\ref{condition poles finite im}) is the usual condition of phase matching after one full cycle.
In the limit of $L\rightarrow \infty$, these equations reduce to 
\begin{align}
	\real[\kappa] &= 0
	\label{condition poles re}\\
	\img[\kappa] &= a, 
	\label{condition poles im}
\end{align}
where $a$ is an arbitrary real number because ${n \over L}\pi$ becomes a dense set. In this case equation \eqref{condition poles im} becomes  immaterial as there is always some $a$ that satisfies \eqref{condition poles im}. 

We also verified the appearance of poles in $r_{\rm s}$ by solving the reflection problem \eqref{reflection slab finite} and \eqref{EB ref finite} numerically. 
In Fig.~\ref{fig Rfinite}, we show numerical results for $r_{\rm s}$ for slabs with thicknesses $L=3\times10^{-3}$m and $9\times 10^{-4}$m respectively. 
Poles correspond to the small white dots in these figures. 	

A useful way of understanding the solutions in the limit of large $L$ is to consider solutions of the equation $	\real[\kappa] = 0$ as defining lines in the complex $\omega$ plane. By imposing additional condition \eqref{condition poles finite im} we find discreet points corresponding to polariton eigenmodes. The distance between solutions obtained using this procedure scales as $O(1/L)$ in agreement with our intuition for the mode quantization in a cavity of length $L$. 
In our numerical calculation, the solution of \eqref{condition poles finite im} we have eigenmode close to the real $\omega$ axis, that correspond to the usual standing wave solutions. We also find solutions that extend into the complex $\omega$ plane, that are special to the Floquet system. These solutions extend between $-1.7{\rm THz}<\img[\omega_{\rm s}]<1.7x{\rm THz}$. As we increase the thickness of the slab, the poles become denser, and in the limit of infinite slab width we find lines of discontinuities of the reflection coefficient.


\section{Angular dependence of reflectivity}\label{Subsec nonnormal}

In this section we consider the problem of light reflection from the Floquet insulator
described by equations (\ref{Maxwell matrix}) for a finite angle of incidence. The goal of this analysis is two-fold. Firstly understanding of the angular dependence of the reflection coefficient is important for analyzing results of light reflection the probe part of the experiments. Secondly, when discussing lasing instabilities in Section III C we only considered modes that have zero momentum in the direction parallel to the surface of the material. We expect
that unstable modes with finite in-plane component are also present in the system.  We remind the readers that our method of analyzing lasing instabilities is to study poles of the reflection coefficient. When incident light has finite momentum along the surface of the sample, the component of momentum parallel to the interface is conserved upon reflection. Hence by analyzing poles of the reflection coefficient at finite angle of incidence, we obtain information about lasing modes with a finite parallel component of momentum. Experimentally they correspond to modes radiating at finite angle to the normal of the sample. 

Following the usual convention for analyzing reflection problems of electro-magnetic waves we discuss separately the cases of s- and p-polarization of light. In the case of s-polarization electric field is parallel to the surface of the material for the incident wave as well as for the reflected and transmitted waves. In the case of p-polarization, magnetic field is parallel to the interface for the incident, scattered, and reflected waves. In this section we present results for the reflection coefficients and focus on the discussion of their physical implications. Details of the derivation can be found in section \ref{Subsec nonnormal derivation}.

\begin{figure}[htp]
	\begin{center}
		\includegraphics[width=0.4\textwidth]{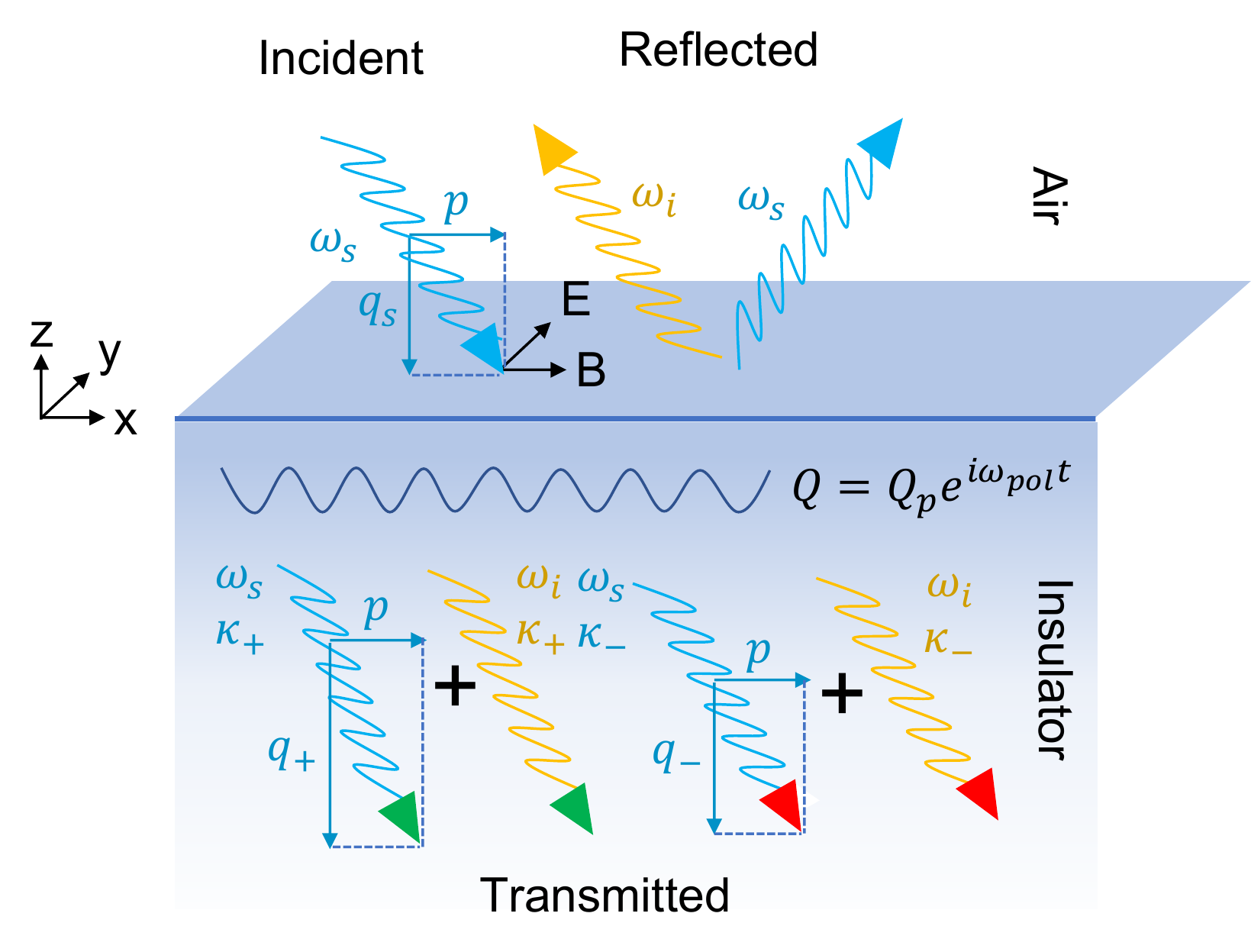}
		\caption{Schematic figure of the Fresnel-Floquet light reflection. The incident light has a frequency $\omega_{\rm s}$ while the reflected light has components of frequencies $\omega_{\rm s}$ and $\omega_{\rm i}$ due to the parametric amplification process. Inside the slab, large amplitude of phonon mixes them; the Floquet eigenmodes are the superpositions of these two components which take the same spatial dependence. }
		\label{fig fresnel nonnormal}
	\end{center}
\end{figure}

To analyze reflection of the S-polarized wave, it is convenient to formulate Maxwell equations using the electric field. We separate the signal and the idler components of the 
electric field and write
\begin{align}
{\bf E}_{\rm s}({\bf r})
= \begin{cases}
E_0 \hat{\bf y} e^{px} (e^{q_{\rm s} z}+r_{\rm s} e^{-q_{\rm s} z})& (0<z) \\
E_0 \hat{\bf y} e^{px} (t_{s+}e^{q_+ z}+t_{s-} e^{q_- z})& (z<0)  \end{cases} \label{Es angled main}\\
{\bf E}_{\rm i}^*({\bf r})
= \begin{cases}
E_0 \hat{\bf y} e^{px} (r_{\rm i} e^{-q_{\rm i} z})& (0<z) \\
E_0 \hat{\bf y} e^{px} (t_{i+}e^{q_+ z}+t_{i-} e^{q_- z})& (z<0)  \end{cases} \label{Ei angled main}
\end{align}
Here $p, q_{\rm s},q_{\rm i}, q_{\pm}$ are imaginary wavevectors that satisfy
\begin{align}
&p^2+q_{\rm s}^2= -{\omega_{\rm s}^2 \over c^2}   \label{p s main}\\
&p^2+q_{\rm i}^2= -{\omega_{\rm i}^2 \over c^2}   \\
&p^2+q_{\pm}^2 = \kappa_\pm^2,
\label{p pm main}
\end{align}
and $r_{\rm s}, r_{\rm i}$ and $t_{s+},t_{s-},t_{i+},t_{i-}$ are the reflection and transmission coefficients, respectively. 
These notations are summarized in Fig.~\ref{fig fresnel nonnormal}. 
The boundary conditions for the Maxwell equations require that the in-plane component of the wave vector to be the same in the air and inside the medium. Note that in equations \eqref{Es angled main} and \eqref{Ei angled main} $p$ as purely imaginary, corresponding to real in-plane momentum. We find
\begin{align}
r_{\rm S} 
= {
	\alpha_-(q_{\rm i}+q_-)(q_{\rm s}-q_+)-\alpha_+(q_{\rm i}+q_+)(q_{\rm s}-q_-)
	\over
	\alpha_-(q_{\rm i}+q_-)(q_{\rm s}+q_+)-\alpha_+(q_{\rm i}+q_+)(q_{\rm s}+q_-)	
}. \label{rs spol}
\end{align}
For the P-polarized wave, we start with the magnetic field. 
\begin{align}
{\bf B}_{\rm s}({\bf r})
= \begin{cases}
B_0 \hat{\bf y} e^{px} (e^{q_{\rm s} z}+r_{\rm s} e^{-q_{\rm s} z})& (0<z) \\
B_0 \hat{\bf y} e^{px} (t_{s+}e^{q_+ z}+t_{s-} e^{q_- z})& (z<0)  \end{cases} \label{Bs p-angled}\\
{\bf B}_{\rm i}^*({\bf r})
= \begin{cases}
B_0 \hat{\bf y} e^{px} (r_{\rm i} e^{-q_{\rm i} z})& (0<z) \\
B_0 \hat{\bf y} e^{px} (t_{i+}e^{q_+ z}+t_{i-} e^{q_- z})& (z<0)  \end{cases} \label{Bi p-angled}
\end{align}
Analysis presented in section (\ref{Subsec nonnormal derivation}) shows that in this case the reflectivity is given by
\begin{align}
r_{\rm P}
= {
	\alpha_-({q_{\rm i}\over \kappa_{\rm i}^2}+{q_- \over \kappa_-^2})({q_{\rm s} \over \kappa_{\rm s}^2}-{q_+ \over \kappa_+^2})-\alpha_+({q_{\rm i}\over \kappa_{\rm i}^2}+{q_+ \over \kappa_+^2})({q_{\rm s} \over \kappa_{\rm s}^2}-{q_- \over \kappa_-^2})
	\over
	\alpha_-({q_{\rm i}\over \kappa_{\rm i}^2}+{q_- \over \kappa_-^2})({q_{\rm s} \over \kappa_{\rm s}^2}+{q_+ \over \kappa_+^2})-\alpha_+({q_{\rm i}\over \kappa_{\rm i}^2}+{q_+ \over \kappa_+^2})({q_{\rm s} \over \kappa_{\rm s}^2}+{q_- \over \kappa_-^2})	
}. \label{rs ppol}
\end{align}

To understand the effect of finite $p$ in equations (\ref{rs spol}) and (\ref{rs ppol})
we recall that discontinuities of the reflection coefficient and the corresponding lasing modes appear when $ \img [\kappa^2]$ crosses zero and $ \real [\kappa^2] <0 $. The latter condition is important since for $\real [\kappa^2] \geq 0$ there is no discontinuity of $\img \kappa$. If we examine equation (\ref{p pm main}) and recall that $p^2$ is negative, we realize that $\real [\kappa^2] -p^2$ can become positive when $|p|$ is sufficiently large. We also recall that we defined the unstable mode to be $\kappa_+$ and that at the lasing transition $ \img [\kappa^2]$ becomes equal to zero.
Hence the lasing instability becomes suppressed when 
\begin{align}
|p| > \kappa_+(\omega^*).
\label{pc_condition}
\end{align}
where $\omega^*$ is the corresponding instability frequency in the bulk ( we remind the readers that instabilities at 20 and 39 THz correspond to the same unstable Floquet eigenmode and for both of them we should take $\omega^*=20$ THz). When condition (\ref{pc_condition}) is not satisfied, we find that instabilities persist although the jump in $q_\pm$ becomes suppressed. This explains why discontinuities in $r_s$ at 20 and 39 THz
become suppressed with the increasing angle of incidence (see Fig. \ref{fig nonnormal}). 
This suggests that lasing instability exists for all modes that have $|p|$ smaller than the critical value of $\frac{\omega_*}{c} \sqrt{\epsilon_0}$, beyond which we can no longer find transverse momentum to match the resonant scattering process shown in Fig~\ref{fig schematic}. For the unstable mode at  $\omega^*=29$ THz we observe that $Re [\kappa^2]$ is small, hence this instability gets suppressed at small values of $|p|$.

\section{Reflection problem for finite angle of incidence} 
\label{Subsec nonnormal derivation}
We present the details of Appendix~\ref{Subsec nonnormal}. 
When the incident wave is not perpendicular to the surface, the reflection and transmission coefficients depend on the polarization of the incident light. A common approach is to decompose incoming waves into S- and P- polarization components. 
For the S-polarized light the electric field is given by \eqref{Es angled main} and \eqref{Ei angled main}. 
We take the direction of the electric field parallel to $y$-axis. 
We can find expressions for the magnetic field using equation \eqref{Maxwell E} and relating it to ${\bf E}_{\rm s}({\bf r})$ and ${\bf E}_{\rm i}^*({\bf r})$
\begin{widetext} 
\begin{align}
{\bf B}_{\rm s}({\bf r})
= \begin{cases}
    -{E_0 \over i \omega_{\rm s}} e^{px} \left[e^{q_{\rm s} z}(q_{\rm s} \hat{\bf x}+p\hat{\bf z}) +r_{\rm s} e^{-q_{\rm s} z}(-q_{\rm s} \hat{\bf x}+p\hat{\bf z}) \right]& (0<z) \\
    -{E_0 \over i \omega_{\rm s}} e^{px} \left[t_{s+}e^{q_+ z}(q_+ \hat{\bf x}+p\hat{\bf z}) +t_{s-} e^{q_- z}(q_- \hat{\bf x}+p\hat{\bf z}) \right]& (z<0) \\
    \end{cases} \label{Bs angled}\\
{\bf B}^*_{\rm i}({\bf r})
= \begin{cases}
    -{E_0 \over i \omega_{\rm i}} e^{px} r_{\rm i} e^{-q_{\rm i} z}(-q_{\rm i} \hat{\bf x}+p\hat{\bf z}) & (0<z) \\
    -{E_0 \over i \omega_{\rm i}} e^{px} \left[t_{i+}e^{q_+ z}(q_+ \hat{\bf x}+p\hat{\bf z}) +t_{i-} e^{q_- z}(q_- \hat{\bf x}+p\hat{\bf z}) \right]& (z<0) \\
    \end{cases} \label{Bi angled}
\end{align}
\end{widetext}
The tangential component of the electric field and the magnetic field should be continuous at $z=0$. 
Therefore, from equations \eqref{Es angled main}, \eqref{Ei angled main}, \eqref{Bs angled}, and \eqref{Bi angled}, we find
\begin{align}
	&1+r_{\rm s} = t_{s+} + t_{s-} \label{boundary 1}\\
	&r_{\rm i} = t_{i+} + t_{i-} \label{boundary 2}\\
	&q_{\rm s}(1-r_{\rm s})= q_{+}t_{s+} + q_{-}t_{s-} \label{boundary 3}\\
	&q_{i}r_{\rm i} = q_{+}t_{s+} + q_{-}t_{s-}.\label{boundary 4}
\end{align} 
Electromagnetic eigenmodes inside the slab are solutions of the Floquet problem discussed in Section II C, hence $t_{s\pm}$ and $t_{i\pm}$ satisfy relation \eqref{relation Es Ei}. Substituting \eqref{Es angled main} and \eqref{Ei angled main} into \eqref{relation Es Ei}, we obtain
\begin{align}
	t_{i\pm}=\alpha_\pm t_{s\pm} \label{relation ti ts}
\end{align}

Solving \eqref{boundary 1}-\eqref{boundary 4} and \eqref{relation ti ts}, the reflection and transmission coefficients are given by
\begin{align}
	r_{\rm s} 
	= {
	\alpha_-(q_{\rm i}+q_-)(q_{\rm s}-q_+)-\alpha_+(q_{\rm i}+q_+)(q_{\rm s}-q_-)
	\over
	\alpha_-(q_{\rm i}+q_-)(q_{\rm s}+q_+)-\alpha_+(q_{\rm i}+q_+)(q_{\rm s}+q_-)	
	}\\
	r_{\rm i}
	= {
	2\alpha_-\alpha_+ q_{\rm s}(-q_++q_-)
	\over
	\alpha_-(q_{\rm i}+q_-)(q_{\rm s}+q_+)-\alpha_+(q_{\rm i}+q_+)(q_{\rm s}+q_-)	
	}\\
	t_{s+}
	= {
	2\alpha_- q_{\rm s}(q_{\rm i}+q_-)
	\over
	\alpha_-(q_{\rm i}+q_-)(q_{\rm s}+q_+)-\alpha_+(q_{\rm i}+q_+)(q_{\rm s}+q_-)	
	}\\
	t_{s-}
	= {
	-2\alpha_+ q_{\rm s}(q_{\rm i}+q_+)
	\over
	\alpha_-(q_{\rm i}+q_-)(q_{\rm s}+q_+)-\alpha_+(q_{\rm i}+q_+)(q_{\rm s}+q_-)	
	}\\
	t_{i+}
	= {
	2\alpha_+\alpha_- q_{\rm s}(q_{\rm i}+q_-)
	\over
	\alpha_-(q_{\rm i}+q_-)(q_{\rm s}+q_+)-\alpha_+(q_{\rm i}+q_+)(q_{\rm s}+q_-)	
	}\\
	t_{i-}
	= {
	-2\alpha_+ \alpha_-q_{\rm s}(q_{\rm i}+q_+)
	\over
	\alpha_-(q_{\rm i}+q_-)(q_{\rm s}+q_+)-\alpha_+(q_{\rm i}+q_+)(q_{\rm s}+q_-)	
	}
\end{align}

The reflection problem is solved for the P-polarized incident light in the same way. 
We can use equation \eqref{Maxwell E} to find the electric fields from the expressions for
 ${\bf B}_{\rm s}({\bf r})$ and ${\bf B}_{\rm i}^*({\bf r})$ in \eqref{Bs p-angled} and \eqref{Bi p-angled}
\begin{widetext} 
\begin{align}
&{\bf E}_{\rm s}({\bf r})
= \begin{cases}
    -{cB_0\over i \omega_{\rm s}}  e^{px} \left[{1\over \epsilon_0}e^{q_{\rm s} z}(q_{\rm s} \hat{\bf x}+p\hat{\bf z}) +{r_{\rm s}\over \epsilon_0} e^{-q_{\rm s} z}(-q_{\rm s} \hat{\bf x}+p\hat{\bf z}) \right]& (0<z) \\
    -{B_0 \over i \omega_{\rm s}} e^{px} \left[{t_{s+}\over \tilde{\epsilon}(\omega_{\rm s})}e^{q_+ z}(q_+ \hat{\bf x}+p\hat{\bf z}) 
    	+{t_{s-}\over \tilde{\epsilon}(\omega_{\rm s})} e^{q_- z}(q_- \hat{\bf x}+p\hat{\bf z}) \right]& (z<0) \\
    \end{cases} \label{Es p-angled}\\
&{\bf E}^*_{\rm i}({\bf r})
= \begin{cases}
    -{cB_0 \over i \omega_{\rm i}} e^{px} {r_{\rm i}\over \epsilon_0} e^{-q_{\rm i} z}(-q_{\rm i} \hat{\bf x}+p\hat{\bf z}) & (0<z) \\
    -{cB_0 \over i \omega_{\rm i}} e^{px} \left[{t_{i+}\over \tilde{\epsilon}(-\omega_{\rm i})}e^{q_+ z}(q_+ \hat{\bf x}+p\hat{\bf z}) +{t_{i-}\over \tilde{\epsilon}(-\omega_{\rm i})} e^{q_- z}(q_- \hat{\bf x}+p\hat{\bf z}) \right]& (z<0) \\
    \end{cases} \label{Ei p-angled}
\end{align}
\end{widetext}
Boundary conditions for the tangential component of electric field and the magnetic field at $z=0$ give
\begin{align}
	&1+r_{\rm s} = t_{s+} + t_{s-} \label{boundary p1}\\
	&r_{\rm i} = t_{i+} + t_{i-} \label{boundary p2}\\
	&{q_{\rm s} \over \epsilon_0}(1-r_{\rm s})= {q_{+}\over \tilde{\epsilon}(\omega_{\rm s})}t_{s+} + {q_{-}\over \tilde{\epsilon}(\omega_{\rm s})}t_{s-} \label{boundary p3}\\
	&{q_{i}\over \epsilon_0}r_{\rm i} = {q_{+}\over \tilde{\epsilon}(-\omega_{\rm i})}t_{s+} + {q_{-}\over \tilde{\epsilon}(-\omega_{\rm i})}t_{s-}.\label{boundary p4}
\end{align}
Solving \eqref{relation ti ts} and \eqref{boundary p1}-\eqref{boundary p4}, the reflection and transmission coefficients are given by
\begin{align}
	r_{\rm s} 
	= {
	\alpha_-({q_{\rm i}\over \kappa_{\rm i}^2}+{q_- \over \kappa_-^2})({q_{\rm s} \over \kappa_{\rm s}^2}-{q_+ \over \kappa_+^2})-\alpha_+({q_{\rm i}\over \kappa_{\rm i}^2}+{q_+ \over \kappa_+^2})({q_{\rm s} \over \kappa_{\rm s}^2}-{q_- \over \kappa_-^2})
	\over
	\alpha_-({q_{\rm i}\over \kappa_{\rm i}^2}+{q_- \over \kappa_-^2})({q_{\rm s} \over \kappa_{\rm s}^2}+{q_+ \over \kappa_+^2})-\alpha_+({q_{\rm i}\over \kappa_{\rm i}^2}+{q_+ \over \kappa_+^2})({q_{\rm s} \over \kappa_{\rm s}^2}+{q_- \over \kappa_-^2})	
	}\\
	r_{\rm i}
	= {
	2\alpha_-\alpha_+ {q_{\rm s} \over \kappa_{\rm s}^2}(-{q_+ \over \kappa_+^2}+{q_- \over \kappa_-^2})
	\over
	\alpha_-({q_{\rm i}\over \kappa_{\rm i}^2}+{q_- \over \kappa_-^2})({q_{\rm s} \over \kappa_{\rm s}^2}+{q_+ \over \kappa_+^2})-\alpha_+({q_{\rm i}\over \kappa_{\rm i}^2}+{q_+ \over \kappa_+^2})({q_{\rm s} \over \kappa_{\rm s}^2}+{q_- \over \kappa_-^2})
		}\\
	t_{s+}
	= {
	2\alpha_- {q_{\rm s} \over \kappa_{\rm s}^2}({q_{\rm i}\over \kappa_{\rm i}^2}+{q_- \over \kappa_-^2})
	\over
	\alpha_-({q_{\rm i}\over \kappa_{\rm i}^2}+{q_- \over \kappa_-^2})({q_{\rm s} \over \kappa_{\rm s}^2}+{q_+ \over \kappa_+^2})-\alpha_+({q_{\rm i}\over \kappa_{\rm i}^2}+{q_+ \over \kappa_+^2})({q_{\rm s} \over \kappa_{\rm s}^2}+{q_- \over \kappa_-^2})
	}\\
	t_{s-}
	= {
	-2\alpha_+ {q_{\rm s} \over \kappa_{\rm s}^2}({q_{\rm i}\over \kappa_{\rm i}^2}+{q_+ \over \kappa_+^2})
	\over
	\alpha_-({q_{\rm i}\over \kappa_{\rm i}^2}+{q_- \over \kappa_-^2})({q_{\rm s} \over \kappa_{\rm s}^2}+{q_+ \over \kappa_+^2})-\alpha_+({q_{\rm i}\over \kappa_{\rm i}^2}+{q_+ \over \kappa_+^2})({q_{\rm s} \over \kappa_{\rm s}^2}+{q_- \over \kappa_-^2})	
	}\\
	t_{i+}
	= {
	2\alpha_+\alpha_- {q_{\rm s} \over \kappa_{\rm s}^2}({q_{\rm i}\over \kappa_{\rm i}^2}+{q_- \over \kappa_-^2})
	\over
	\alpha_-({q_{\rm i}\over \kappa_{\rm i}^2}+{q_- \over \kappa_-^2})({q_{\rm s} \over \kappa_{\rm s}^2}+{q_+ \over \kappa_+^2})-\alpha_+({q_{\rm i}\over \kappa_{\rm i}^2}+{q_+ \over \kappa_+^2})({q_{\rm s} \over \kappa_{\rm s}^2}+{q_- \over \kappa_-^2})
		}\\
	t_{i-}
	= {
	-2\alpha_+ \alpha_-{q_{\rm s} \over \kappa_{\rm s}^2}({q_{\rm i}\over \kappa_{\rm i}^2}+{q_+ \over \kappa_-^2})
	\over
	\alpha_-({q_{\rm i}\over \kappa_{\rm i}^2}+{q_- \over \kappa_-^2})({q_{\rm s} \over \kappa_{\rm s}^2}+{q_+ \over \kappa_+^2})-\alpha_+({q_{\rm i}\over \kappa_{\rm i}^2}+{q_+ \over \kappa_+^2})({q_{\rm s} \over \kappa_{\rm s}^2}+{q_- \over \kappa_-^2})
		}
\end{align}
We note that we ignore cubic terms in the effective potential \eqref{Hamiltonian nonlinear} here. In the P-polarized case, the effective potential of 4H-SiC has cubic terms in general. We should take these terms into account for more accurate analysis.

\bibliography{phonon_laser_arXiv.bib}

\begin{thebibliography}{20}%
\makeatletter
\providecommand \@ifxundefined [1]{%
 \@ifx{#1\undefined}
}%
\providecommand \@ifnum [1]{%
 \ifnum #1\expandafter \@firstoftwo
 \else \expandafter \@secondoftwo
 \fi
}%
\providecommand \@ifx [1]{%
 \ifx #1\expandafter \@firstoftwo
 \else \expandafter \@secondoftwo
 \fi
}%
\providecommand \natexlab [1]{#1}%
\providecommand \enquote  [1]{``#1''}%
\providecommand \bibnamefont  [1]{#1}%
\providecommand \bibfnamefont [1]{#1}%
\providecommand \citenamefont [1]{#1}%
\providecommand \href@noop [0]{\@secondoftwo}%
\providecommand \href [0]{\begingroup \@sanitize@url \@href}%
\providecommand \@href[1]{\@@startlink{#1}\@@href}%
\providecommand \@@href[1]{\endgroup#1\@@endlink}%
\providecommand \@sanitize@url [0]{\catcode `\\12\catcode `\$12\catcode `\&12\catcode `\#12\catcode `\^12\catcode `\_12\catcode `\%12\relax}%
\providecommand \@@startlink[1]{}%
\providecommand \@@endlink[0]{}%
\providecommand \url  [0]{\begingroup\@sanitize@url \@url }%
\providecommand \@url [1]{\endgroup\@href {#1}{\urlprefix }}%
\providecommand \urlprefix  [0]{URL }%
\providecommand \Eprint [0]{\href }%
\providecommand \doibase [0]{http://dx.doi.org/}%
\providecommand \selectlanguage [0]{\@gobble}%
\providecommand \bibinfo  [0]{\@secondoftwo}%
\providecommand \bibfield  [0]{\@secondoftwo}%
\providecommand \translation [1]{[#1]}%
\providecommand \BibitemOpen [0]{}%
\providecommand \bibitemStop [0]{}%
\providecommand \bibitemNoStop [0]{.\EOS\space}%
\providecommand \EOS [0]{\spacefactor3000\relax}%
\providecommand \BibitemShut  [1]{\csname bibitem#1\endcsname}%
\let\auto@bib@innerbib\@empty
\bibitem [{\citenamefont {Cartella}\ \emph {et~al.}(2018)\citenamefont {Cartella}, \citenamefont {Nova}, \citenamefont {Fechner}, \citenamefont {Merlin},\ and\ \citenamefont {Cavalleri}}]{Cartella12148}%
  \BibitemOpen
  \bibfield  {author} {\bibinfo {author} {\bibfnamefont {A.}~\bibnamefont {Cartella}}, \bibinfo {author} {\bibfnamefont {T.~F.}\ \bibnamefont {Nova}}, \bibinfo {author} {\bibfnamefont {M.}~\bibnamefont {Fechner}}, \bibinfo {author} {\bibfnamefont {R.}~\bibnamefont {Merlin}}, \ and\ \bibinfo {author} {\bibfnamefont {A.}~\bibnamefont {Cavalleri}},\ }\href {\doibase 10.1073/pnas.1809725115} {\bibfield  {journal} {\bibinfo  {journal} {Proceedings of the National Academy of Sciences}\ }\textbf {\bibinfo {volume} {115}},\ \bibinfo {pages} {12148} (\bibinfo {year} {2018})},\ \Eprint {http://arxiv.org/abs/https://www.pnas.org/content/115/48/12148.full.pdf} {https://www.pnas.org/content/115/48/12148.full.pdf} \BibitemShut {NoStop}%
\bibitem [{\citenamefont {Leitenstorfer}\ \emph {et~al.}(1999)\citenamefont {Leitenstorfer}, \citenamefont {Hunsche}, \citenamefont {Shah}, \citenamefont {Nuss},\ and\ \citenamefont {Knox}}]{LeitenStorferPRL}%
  \BibitemOpen
  \bibfield  {author} {\bibinfo {author} {\bibfnamefont {A.}~\bibnamefont {Leitenstorfer}}, \bibinfo {author} {\bibfnamefont {S.}~\bibnamefont {Hunsche}}, \bibinfo {author} {\bibfnamefont {J.}~\bibnamefont {Shah}}, \bibinfo {author} {\bibfnamefont {M.~C.}\ \bibnamefont {Nuss}}, \ and\ \bibinfo {author} {\bibfnamefont {W.~H.}\ \bibnamefont {Knox}},\ }\href {\doibase 10.1103/PhysRevLett.82.5140} {\bibfield  {journal} {\bibinfo  {journal} {Phys. Rev. Lett.}\ }\textbf {\bibinfo {volume} {82}},\ \bibinfo {pages} {5140} (\bibinfo {year} {1999})}\BibitemShut {NoStop}%
\bibitem [{\citenamefont {Leitenstorfer}\ \emph {et~al.}(2000)\citenamefont {Leitenstorfer}, \citenamefont {Hunsche}, \citenamefont {Shah}, \citenamefont {Nuss},\ and\ \citenamefont {Knox}}]{LeitenstorferPRB}%
  \BibitemOpen
  \bibfield  {author} {\bibinfo {author} {\bibfnamefont {A.}~\bibnamefont {Leitenstorfer}}, \bibinfo {author} {\bibfnamefont {S.}~\bibnamefont {Hunsche}}, \bibinfo {author} {\bibfnamefont {J.}~\bibnamefont {Shah}}, \bibinfo {author} {\bibfnamefont {M.~C.}\ \bibnamefont {Nuss}}, \ and\ \bibinfo {author} {\bibfnamefont {W.~H.}\ \bibnamefont {Knox}},\ }\href {\doibase 10.1103/PhysRevB.61.16642} {\bibfield  {journal} {\bibinfo  {journal} {Phys. Rev. B}\ }\textbf {\bibinfo {volume} {61}},\ \bibinfo {pages} {16642} (\bibinfo {year} {2000})}\BibitemShut {NoStop}%
\bibitem [{\citenamefont {Skaar}(2006)}]{Johannes2006}%
  \BibitemOpen
  \bibfield  {author} {\bibinfo {author} {\bibfnamefont {J.}~\bibnamefont {Skaar}},\ }\href {\doibase 10.1103/PhysRevE.73.026605} {\bibfield  {journal} {\bibinfo  {journal} {Phys. Rev. E}\ }\textbf {\bibinfo {volume} {73}},\ \bibinfo {pages} {026605} (\bibinfo {year} {2006})}\BibitemShut {NoStop}%
\bibitem [{\citenamefont {Dorofeenko}\ \emph {et~al.}(2012)\citenamefont {Dorofeenko}, \citenamefont {Zyablovsky}, \citenamefont {Pukhov}, \citenamefont {Lisyansky},\ and\ \citenamefont {Vinogradov}}]{Dorofeenko2012}%
  \BibitemOpen
  \bibfield  {author} {\bibinfo {author} {\bibfnamefont {A.~V.}\ \bibnamefont {Dorofeenko}}, \bibinfo {author} {\bibfnamefont {A.~A.}\ \bibnamefont {Zyablovsky}}, \bibinfo {author} {\bibfnamefont {A.~A.}\ \bibnamefont {Pukhov}}, \bibinfo {author} {\bibfnamefont {A.~A.}\ \bibnamefont {Lisyansky}}, \ and\ \bibinfo {author} {\bibfnamefont {A.~P.}\ \bibnamefont {Vinogradov}},\ }\href {\doibase 10.3367/ufne.0182.201211b.1157} {\bibfield  {journal} {\bibinfo  {journal} {Physics-Uspekhi}\ }\textbf {\bibinfo {volume} {55}},\ \bibinfo {pages} {1080} (\bibinfo {year} {2012})}\BibitemShut {NoStop}%
\bibitem [{\citenamefont {Carusotto}\ and\ \citenamefont {Ciuti.}(2013)}]{carusotto2013}%
  \BibitemOpen
  \bibfield  {author} {\bibinfo {author} {\bibfnamefont {I.}~\bibnamefont {Carusotto}}\ and\ \bibinfo {author} {\bibfnamefont {C.}~\bibnamefont {Ciuti.}},\ }\href@noop {} {\bibfield  {journal} {\bibinfo  {journal} {Rev. Mod. Phys.}\ }\textbf {\bibinfo {volume} {85}},\ \bibinfo {pages} {299} (\bibinfo {year} {2013})}\BibitemShut {NoStop}%
\bibitem [{\citenamefont {edited~by R.~A.~Fisher}(1983)}]{Fisher}%
  \BibitemOpen
  \bibfield  {author} {\bibinfo {author} {\bibnamefont {edited~by R.~A.~Fisher}},\ }\href@noop {} {\emph {\bibinfo {title} {Optical Phase Conjugation}}}\ (\bibinfo  {publisher} {Elsevier},\ \bibinfo {year} {1983})\BibitemShut {NoStop}%
\bibitem [{\citenamefont {Zeldovich}\ \emph {et~al.}(1985)\citenamefont {Zeldovich}, \citenamefont {Piptesky},\ and\ \citenamefont {Shkunov}}]{Zeldovich}%
  \BibitemOpen
  \bibfield  {author} {\bibinfo {author} {\bibnamefont {Zeldovich}}, \bibinfo {author} {\bibnamefont {Piptesky}}, \ and\ \bibinfo {author} {\bibnamefont {Shkunov}},\ }\href@noop {} {\emph {\bibinfo {title} {Principles of Phase Conjugation}}}\ (\bibinfo  {publisher} {Springer},\ \bibinfo {year} {1985})\BibitemShut {NoStop}%
\bibitem [{\citenamefont {Yariv}(1989)}]{Yariv}%
  \BibitemOpen
  \bibfield  {author} {\bibinfo {author} {\bibfnamefont {A.}~\bibnamefont {Yariv}},\ }\href@noop {} {\emph {\bibinfo {title} {Quantum Electronics}}}\ (\bibinfo  {publisher} {John Wiley \& Sons},\ \bibinfo {year} {1989})\BibitemShut {NoStop}%
\bibitem [{\citenamefont {Blonder}\ \emph {et~al.}(1982)\citenamefont {Blonder}, \citenamefont {Tinkham},\ and\ \citenamefont {Klapwijk}}]{Tinkham}%
  \BibitemOpen
  \bibfield  {author} {\bibinfo {author} {\bibfnamefont {G.~E.}\ \bibnamefont {Blonder}}, \bibinfo {author} {\bibnamefont {Tinkham}}, \ and\ \bibinfo {author} {\bibfnamefont {T.~M.}\ \bibnamefont {Klapwijk}},\ }\href@noop {} {\bibfield  {journal} {\bibinfo  {journal} {Phys. Rev. B}\ }\textbf {\bibinfo {volume} {25}},\ \bibinfo {pages} {4515} (\bibinfo {year} {1982})}\BibitemShut {NoStop}%
\bibitem [{\citenamefont {Buzzi}\ \emph {et~al.}(2019)\citenamefont {Buzzi}, \citenamefont {Jotzu}, \citenamefont {Cavalleri}, \citenamefont {Cirac}, \citenamefont {Demler}, \citenamefont {Halperin}, \citenamefont {Lukin}, \citenamefont {Shi}, \citenamefont {Wang},\ and\ \citenamefont {Podolsky}}]{Buzzi2019}%
  \BibitemOpen
  \bibfield  {author} {\bibinfo {author} {\bibfnamefont {M.}~\bibnamefont {Buzzi}}, \bibinfo {author} {\bibfnamefont {G.}~\bibnamefont {Jotzu}}, \bibinfo {author} {\bibfnamefont {A.}~\bibnamefont {Cavalleri}}, \bibinfo {author} {\bibfnamefont {J.~I.}\ \bibnamefont {Cirac}}, \bibinfo {author} {\bibfnamefont {E.~A.}\ \bibnamefont {Demler}}, \bibinfo {author} {\bibfnamefont {B.~I.}\ \bibnamefont {Halperin}}, \bibinfo {author} {\bibfnamefont {M.~D.}\ \bibnamefont {Lukin}}, \bibinfo {author} {\bibfnamefont {T.}~\bibnamefont {Shi}}, \bibinfo {author} {\bibfnamefont {Y.}~\bibnamefont {Wang}}, \ and\ \bibinfo {author} {\bibfnamefont {D.}~\bibnamefont {Podolsky}},\ }\href@noop {} {\bibfield  {journal} {\bibinfo  {journal} {arXiv 1908.10879}\ } (\bibinfo {year} {2019})}\BibitemShut {NoStop}%
\bibitem [{\citenamefont {Sivak}\ \emph {et~al.}(2019)\citenamefont {Sivak}, \citenamefont {Frattini}, \citenamefont {Joshi}, \citenamefont {Lingenfelter}, \citenamefont {Shankar},\ and\ \citenamefont {Devoret}}]{PhysRevApplied.11.054060}%
  \BibitemOpen
  \bibfield  {author} {\bibinfo {author} {\bibfnamefont {V.}~\bibnamefont {Sivak}}, \bibinfo {author} {\bibfnamefont {N.}~\bibnamefont {Frattini}}, \bibinfo {author} {\bibfnamefont {V.}~\bibnamefont {Joshi}}, \bibinfo {author} {\bibfnamefont {A.}~\bibnamefont {Lingenfelter}}, \bibinfo {author} {\bibfnamefont {S.}~\bibnamefont {Shankar}}, \ and\ \bibinfo {author} {\bibfnamefont {M.}~\bibnamefont {Devoret}},\ }\href {\doibase 10.1103/PhysRevApplied.11.054060} {\bibfield  {journal} {\bibinfo  {journal} {Phys. Rev. Applied}\ }\textbf {\bibinfo {volume} {11}},\ \bibinfo {pages} {054060} (\bibinfo {year} {2019})}\BibitemShut {NoStop}%
\bibitem [{\citenamefont {Boyd}(2008)}]{BoydBook}%
  \BibitemOpen
  \bibfield  {author} {\bibinfo {author} {\bibfnamefont {R.}~\bibnamefont {Boyd}},\ }\href@noop {} {\emph {\bibinfo {title} {Nonlinear Optics}}}\ (\bibinfo  {publisher} {Academic Press},\ \bibinfo {year} {2008})\BibitemShut {NoStop}%
\bibitem [{Note1()}]{Note1}%
  \BibitemOpen
  \bibinfo {note} {See { http://pd.chem.ucl.ac.uk/pdnn/symm2/pg6mm.htm}}\BibitemShut {NoStop}%
\bibitem [{Note2()}]{Note2}%
  \BibitemOpen
  \bibinfo {note} {See character table http://www.cryst.ehu.es/cgi-bin/rep/programs/sam/point.py?sg=183\&num=25}\BibitemShut {NoStop}%
\bibitem [{\citenamefont {Drever}\ \emph {et~al.}(1983)\citenamefont {Drever}, \citenamefont {Hall}, \citenamefont {Kowalski}, \citenamefont {Hough}, \citenamefont {Ford}, \citenamefont {Munley},\ and\ \citenamefont {Ward}}]{Drever1983}%
  \BibitemOpen
  \bibfield  {author} {\bibinfo {author} {\bibfnamefont {R.~W.~P.}\ \bibnamefont {Drever}}, \bibinfo {author} {\bibfnamefont {J.~L.}\ \bibnamefont {Hall}}, \bibinfo {author} {\bibfnamefont {F.~V.}\ \bibnamefont {Kowalski}}, \bibinfo {author} {\bibfnamefont {J.}~\bibnamefont {Hough}}, \bibinfo {author} {\bibfnamefont {G.~M.}\ \bibnamefont {Ford}}, \bibinfo {author} {\bibfnamefont {A.~J.}\ \bibnamefont {Munley}}, \ and\ \bibinfo {author} {\bibfnamefont {H.}~\bibnamefont {Ward}},\ }\href {\doibase 10.1007/BF00702605} {\bibfield  {journal} {\bibinfo  {journal} {Applied Physics B}\ }\textbf {\bibinfo {volume} {31}},\ \bibinfo {pages} {97} (\bibinfo {year} {1983})}\BibitemShut {NoStop}%
\bibitem [{\citenamefont {Ying}\ \emph {et~al.}(2007)\citenamefont {Ying}, \citenamefont {Ma},\ and\ \citenamefont {Jin}}]{Ying:07}%
  \BibitemOpen
  \bibfield  {author} {\bibinfo {author} {\bibfnamefont {D.}~\bibnamefont {Ying}}, \bibinfo {author} {\bibfnamefont {H.}~\bibnamefont {Ma}}, \ and\ \bibinfo {author} {\bibfnamefont {Z.}~\bibnamefont {Jin}},\ }\href {\doibase 10.1364/AO.46.004890} {\bibfield  {journal} {\bibinfo  {journal} {Appl. Opt.}\ }\textbf {\bibinfo {volume} {46}},\ \bibinfo {pages} {4890} (\bibinfo {year} {2007})}\BibitemShut {NoStop}%
\bibitem [{\citenamefont {Savvidis}\ \emph {et~al.}(2000)\citenamefont {Savvidis}, \citenamefont {Baumberg}, \citenamefont {Stevenson}, \citenamefont {Skolnick}, \citenamefont {Whittaker},\ and\ \citenamefont {Roberts}}]{Roberts1}%
  \BibitemOpen
  \bibfield  {author} {\bibinfo {author} {\bibfnamefont {P.~G.}\ \bibnamefont {Savvidis}}, \bibinfo {author} {\bibfnamefont {J.~J.}\ \bibnamefont {Baumberg}}, \bibinfo {author} {\bibfnamefont {R.~M.}\ \bibnamefont {Stevenson}}, \bibinfo {author} {\bibfnamefont {M.~S.}\ \bibnamefont {Skolnick}}, \bibinfo {author} {\bibfnamefont {D.~M.}\ \bibnamefont {Whittaker}}, \ and\ \bibinfo {author} {\bibfnamefont {J.~S.}\ \bibnamefont {Roberts}},\ }\href {\doibase 10.1103/PhysRevLett.84.1547} {\bibfield  {journal} {\bibinfo  {journal} {Phys. Rev. Lett.}\ }\textbf {\bibinfo {volume} {84}},\ \bibinfo {pages} {1547} (\bibinfo {year} {2000})}\BibitemShut {NoStop}%
\bibitem [{\citenamefont {Stevenson}\ \emph {et~al.}(2000)\citenamefont {Stevenson}, \citenamefont {Astratov}, \citenamefont {Skolnick}, \citenamefont {Whittaker}, \citenamefont {Emam-Ismail}, \citenamefont {Tartakovskii}, \citenamefont {Savvidis}, \citenamefont {Baumberg},\ and\ \citenamefont {Roberts}}]{Roberts2}%
  \BibitemOpen
  \bibfield  {author} {\bibinfo {author} {\bibfnamefont {R.~M.}\ \bibnamefont {Stevenson}}, \bibinfo {author} {\bibfnamefont {V.~N.}\ \bibnamefont {Astratov}}, \bibinfo {author} {\bibfnamefont {M.~S.}\ \bibnamefont {Skolnick}}, \bibinfo {author} {\bibfnamefont {D.~M.}\ \bibnamefont {Whittaker}}, \bibinfo {author} {\bibfnamefont {M.}~\bibnamefont {Emam-Ismail}}, \bibinfo {author} {\bibfnamefont {A.~I.}\ \bibnamefont {Tartakovskii}}, \bibinfo {author} {\bibfnamefont {P.~G.}\ \bibnamefont {Savvidis}}, \bibinfo {author} {\bibfnamefont {J.~J.}\ \bibnamefont {Baumberg}}, \ and\ \bibinfo {author} {\bibfnamefont {J.~S.}\ \bibnamefont {Roberts}},\ }\href {\doibase 10.1103/PhysRevLett.85.3680} {\bibfield  {journal} {\bibinfo  {journal} {Phys. Rev. Lett.}\ }\textbf {\bibinfo {volume} {85}},\ \bibinfo {pages} {3680} (\bibinfo {year} {2000})}\BibitemShut {NoStop}%
\bibitem [{\citenamefont {Baumberg}\ \emph {et~al.}(2000)\citenamefont {Baumberg}, \citenamefont {Savvidis}, \citenamefont {Stevenson}, \citenamefont {Tartakovskii}, \citenamefont {Skolnick}, \citenamefont {Whittaker},\ and\ \citenamefont {Roberts}}]{Roberts3}%
  \BibitemOpen
  \bibfield  {author} {\bibinfo {author} {\bibfnamefont {J.~J.}\ \bibnamefont {Baumberg}}, \bibinfo {author} {\bibfnamefont {P.~G.}\ \bibnamefont {Savvidis}}, \bibinfo {author} {\bibfnamefont {R.~M.}\ \bibnamefont {Stevenson}}, \bibinfo {author} {\bibfnamefont {A.~I.}\ \bibnamefont {Tartakovskii}}, \bibinfo {author} {\bibfnamefont {M.~S.}\ \bibnamefont {Skolnick}}, \bibinfo {author} {\bibfnamefont {D.~M.}\ \bibnamefont {Whittaker}}, \ and\ \bibinfo {author} {\bibfnamefont {J.~S.}\ \bibnamefont {Roberts}},\ }\href {\doibase 10.1103/PhysRevB.62.R16247} {\bibfield  {journal} {\bibinfo  {journal} {Phys. Rev. B}\ }\textbf {\bibinfo {volume} {62}},\ \bibinfo {pages} {R16247} (\bibinfo {year} {2000})}\BibitemShut {NoStop}%
\end{thebibliography}%
\end{document}